%% file: main.tex
\documentclass[manuscript,screen,authorversion]{acmart}
\renewcommand\footnotetextcopyrightpermission[1]{} 
\usepackage{longtable}
\AtBeginDocument{%
  \providecommand\BibTeX{{%
    \normalfont B\kern-0.5em{\scshape i\kern-0.25em b}\kern-0.8em\TeX}}}

\setcopyright{none}

\begin{document}

\title[Techno-utopians, Scammers, and Bullshitters]{Techno-Utopians, Scammers, and Bullshitters: The Promise and Peril of Web3 and Blockchain Technologies According to Operators and Venture Capital Investors}

\author{Amy A. Winecoff}
\email{aawinecoff@gmail.com}
\affiliation{%
  \institution{Princeton University}
  \city{Princeton}
  \state{NJ}
  \country{USA}
}

\author{Johannes Lenhard}
\email{jfl37@cam.ac.uk}
\affiliation{%
  \institution{University of Cambridge}
  \city{Cambridge}
  \country{UK}
}

\renewcommand{\shortauthors}{Winecoff and Lenhard}

\begin{abstract}
Proponents of Web3 and blockchain argue that these technologies can revolutionize how people live and work by better empowering individuals and distributing influence to a broader base of stakeholders. While technologists often have expansive hopes for what their technologies will accomplish over the long term, the practical challenges of developing, scaling, and maintaining systems amidst present-day constraints can compromise progress toward this vision. How technologists think about the technological future they hope to enable and how they navigate day-to-day issues impacts the form technologies take, their potential benefits, and their potential harms. In our current work, we aimed to explore the visions of Web3 and blockchain technologists and identify the immediate challenges that could threaten their visions. We conducted semi-structured interviews with 29 operators and professional investors in the Web3 and blockchain field. Our findings revealed that participants supported several ideological goals for their projects, with decentralization being a pivotal mechanism to enable user autonomy, distribute governance power, and promote financial inclusion. However, participants acknowledged the practical difficulties in fulfilling these promises, including the need for rapid technology development, conflicts of interest among stakeholders due to platform financing dynamics, and the challenge of expanding to mainstream users who may not share the "Web3 ethos." If negotiated ineffectively, these challenges could lead to negative outcomes, such as corrupt governance, increased inequality, and increased prevalence of scams and dubious investment schemes. While participants thought education, regulation, and a renewed commitment to the original blockchain ideals could alleviate some problems, they expressed skepticism about the potential of these solutions. We conclude by discussing how technologists' pursuit of ideological goals within the Web3 and blockchain sector will continue to shape the impact of these technologies, good and bad.
\end{abstract}





\maketitle

\section{Introduction}

An understanding of the value systems and design principles that animate technology developers can help elucidate the potential benefits and harms of the systems they build. In 2008, the pseudonymous author Satoshi Nakamoto described the design for Bitcoin, a new digital currency leveraging the blockchain that could enable financial transactions without need of centralized oversight or control \cite{nakamoto2008bitcoin}. On a technical level, Bitcoin's design was influenced by the developments of academic computer scientists in the 1980s and 1990s \cite{narayanan2017bitcoin}, but it was also heavily influenced by its cultural context. The social community that gave rise to Bitcoin was comprised of activists, cryptographers, and academics who saw cryptographic techniques as tools to instantiate ideological beliefs, particularly regarding privacy and free markets \cite{maurer2013perhaps, swartz2018bitcoin, hayes2019socio, narayanan2013happened}. For this reason, some have suggested that Bitcoin succeeds as much or more as an ideology than it does as a technology \cite{dodd2018social, knittel2019most, golumbia2016politics}. As Nigel Dodd \cite{dodd2018social} summarizes:  "Bitcoin is arguably a social movement as much as it is a currency." 

Following from this technical and cultural legacy, the applications of blockchain technology have extended to diverse technologies such as non-fungible tokens (NFTs), decentralized autonomous organizations (DAOs), and decentralized finance (DeFi). In their totality, these technologies would form the basis of "Web3," \footnote{We note that Web3 is not the same as Web 3.0, sometimes referred to as the "semantic web" conceptualized by Tim Berners Lee \cite{berners2001semantic}}, a decentralized version of the web that promises to "transform the internet as we know it, upending traditional gatekeepers and ushering in a new, middleman-free digital economy \cite{rooseweb3}." In theory, Web3 would act as a stark contrast to our current version of the web, referred to as "Web2," which is primarily mediated by centralized corporations such as Meta, Google, and Apple. In the words of Gavin Wood, a member of the original team that conceived of the Ethereum blockchain, in Web2, "wealth, power and influence [is] placed in the hands of the greedy, the megalomaniacs, or the plain malicious"\cite{wood2018}. Thus, while the domains of application have multiplied, the developers of today's blockchain and Web3 technologies are still motivated by ideological goals related to decentralization, supplanting human intermediaries with protocols, and user sovereignty  \cite{lustig2019intersecting, semenzin2021automating, husain2020political, brody2021ideologies, inwood2021ideology}.

Like the technologies themselves, the audience of blockchain technology adopters has expanded significantly to include users with hetereogenous motivations and competencies. While a subset of technologically-savvy early adopters report an ideological alignment with decentralization and feel confident in their abilities to safely and successfully navigate blockchain systems, users who are primarily motivated by potential investment opportunities report feeling unable to protect themselves from negative outcomes such as theft of assets \cite{abramova2021bits}. Other literature indicates that many users experience significant usability challenges with blockchain technologies \cite{frohlich2020don}, misunderstand core concepts related to privacy and security within blockchain systems \cite{krombholz2017other}, and often suffer financial losses as a result \cite{krombholz2017other, abramova2021bits}. Thus, technology developers' techno-utopian vision for the future has consequences for current-day users, even if they do not share in the techno-utopian vision.

Revolutionizing social, political, and economic systems is not easy. Along the path toward their long-term goals for their platforms, technologists must wrangle with the practical constraints of building and scaling organizations. In other sectors, founders \cite{winecoff2022artificial} and venture capitalists \cite{lenhard2021} alike frequently compromise their personal values or higher-level missions to deal with practical demands, especially when those demands introduce financial conflicts of interest. At a recent panel on how blockchain companies can address concerns related to environmental, social, and governance (ESG) principles \footnote{VentureESG, Crypto and ESG: How do they work together?, Dec. 2022. bit.ly/3qXdseo}, one venture capital (VC) investor noted: 
\begin{quote}
"There’s a necessary tension between understanding how systems work, and making them better, and making money that is hard to resolve when you do it all at once. [...] the hardest thing for me to resolve as an investor is when to embrace innovation that breaks things, and when to push back."
\end{quote} 

The compromises that technologists make when "innovating and breaking things" also affects these systems' impact on users and society. Within the same panel, panelists acknowledged the sector's apparent problems, such as Bitcoin's environmental footprint, FTX's fraudulent business practices, and the blatant wealth inequality within the ecosystem. Still, they continued to express hope that the sector's higher mission could act as a North Star, guiding innovators through the current market turmoil. 

In our current work, we sought to understand what the technologists developing blockchain technologies viewed as their broader vision for the future of their technologies and what they regard as near-term challenges that threaten this vision. To do so, we conducted semi-structured interviews with 29 operators and professional investors within the Web3 and blockchain space. Our results revealed that participants supported several ideological goals for their projects; decentralization was often cited as a mechanism for supporting user autonomy (i.e., "self-sovereignty"), for distributing governance power to platforms' stakeholder communities, and for supporting financial inclusion. At the same time, participants acknowledged that these promises are difficult to fulfill for practical reasons, such as the need for platform developers to iterate quickly on their technology, platform financing dynamics that create conflicts of interest between different stakeholder groups, and the need to expand to mainstream markets comprised of users who do not necessarily share the "Web3 ethos." If poorly navigated, these challenges could result in perilous outcomes, such as corrupt or incompetent governance, magnification of inequality, and proliferation of scams and dubious investment schemes. Although participants thought education, regulation, and recommitment to the original blockchain ideals could mitigate some problems, they often found more fault than potential in such solutions.  

\section{Related Works}
Prior scholarship on blockchain technology and culture suggests that these technologies were born from their designers' social, political, and economic beliefs, a cultural heritage that continues to shape blockchain technologies today \cite{maurer2013perhaps, swartz2016, swartz2018bitcoin, narayanan2013happened, hayes2019socio, brody2021ideologies, husain2020political, inwood2021ideology}. While much of the rhetoric about the potential of such technologies emphasizes a role for decentralization, researchers have argued that decentralization as a concept is neither binary nor unidimensional. Our study builds on this prior research by examining how investors and operators conceptualize their ideal for the future of technology and the role decentralization plays in it. We briefly review previous literature on these topics to provide context for our own investigation.

\subsection{Ideological Origins of Blockchain \& Web3 Technology}
Sociotechnical research has repeatedly demonstrated that communities' social norms and human values are embedded within the systems they create. Sometimes norms and values are embedded implicitly or unintentionally into systems. For example, machine learning system development implicitly promotes the values of efficiency, universality, and impartiality \cite{scheuerman2021datasets, dotan2019value, birhane2022values}, often to the detriment of other values, such as protecting marginalized people from harms \cite{birhane2022values}. On the other hand, sometimes technologists overtly design systems to promote particular human values through value-sensitive design approaches \cite{friedman2013value, friedman2019value} or through participatory design methods \cite{hope2019hackathons, liaqat2021participatory}. 

Early blockchain systems and the algorithms that comprise them were initially developed by activists, cryptographers, and academics who viewed such systems as a means of achieving ideological goals. Scholars have identified two interacting but distinct subcultures amongst those that pioneered blockchain technology, the cypherpunks and the crypto-anarchists, both of which trace their origin to the Cypherpunk mailing list started in the early 1990s \cite{swartz2018bitcoin, hayes2019socio, narayanan2013happened, swartz2016, maurer2013perhaps}. The central tenet of the cypherpunk ideology and the technologies that arose from it is that privacy is fundamental to free societies. Cypherpunks believed that creating technologies that protect the privacy of individuals from governments and corporations was the ultimate moral act because it enhanced a collective good. Crypto-anarchists extended the notions of freedom grounded in privacy to additional beliefs about unencumbered financial markets. From the crypto-anarchist perspective, a monetary system divorced from state oversight and manipulation was central to establishing a free (market-based) society \cite{swartz2018bitcoin}.

These two notions of freedom--one pertaining to information and expression and the other pertaining to markets--have been linked not only to the design of Bitcoin and the narratives surrounding it \cite{swartz2018bitcoin, maurer2013perhaps, narayanan2013happened, swartz2016}, but also to more recent blockchain developments such as Ethereum \cite{brody2021ideologies}. Scholarly analyses that criticize blockchain systems tend to focus on the free market imaginary. For example, Golumbia \cite{golumbia2016politics} argues that Bitcoin is an outgrowth of right-wing monetary conspiracy theories, and Herian \cite{herian2018blockchain} and Crandall \cite{crandall2019blockchains} argue that blockchain advocacy can be viewed as a neoliberal effort to reinforce the dominance of private companies. More favorable analyses seek to resurface the free expression imaginary of the cypherpunks and the "hacker-engineer" sensibility that seeks to build for the collective good rather than to undermine it \cite{brekke2021hacker}. Over time, the motivations of blockchain advocates have extended beyond either notion of freedom to encompass a broader array of values \cite{husain2020political, semenzin2021automating, inwood2021ideology}. Nevertheless, social, economic, and political beliefs continue to shape the discourse surrounding blockchain technologies and, in some cases, to shape the technologies themselves.

\subsection{Disambiguating Decentralization}
Decentralization tendencies are often identified as occurring at least two levels: the "technical layer" and the "social layer" \cite{lustig2019intersecting, fard2021distributing, walch2019deconstructing}. The technical layer encompasses both the network architecture and the protocols that regulate the participation of nodes within it. Whereas the network architecture defines the topology of the nodes and their connections in the network, the protocols define the rules for how nodes in the network operate \cite{bodo2021decentralisation}. Even systems whose technical designs could theoretically support a decentralized network are subject to centralizing factors. For example, if a large proportion of nodes in a network are located within a single geographic region and that region experiences a natural disaster, the network's topology could change radically over a very short period of time \cite{walch2019deconstructing}. 

The social layer typically refers to how a variety of human stakeholders significantly impact the system's outcome. Stakeholders may include node operators, token holders, miners, core developers, governing bodies (i.e., "foundations"), institutional investors, and so forth. Core developers, the team tasked with maintaining and updating a protocol's codebase, have been pointed to as a source of centralized power within blockchain technology systems because they have the potential to affect the outcome of the system dramatically; this power is further amplified in times of crises such as hacks \cite{dupont2017experiments, hutten2019soft, walch2019deconstructing, shin2022} and major disagreements amongst stakeholders regarding protocol updates \cite{walch2019deconstructing, defillipi2016invisible, caliskan2022rise}. 

Dynamics occurring outside of the system can influence centralization within the technical layer, the social layer, or both. A frequently cited example of this is in mining. Within Bitcoin, to verify transactions and add new blocks to the ledger of transactions, nodes within the network compete to solve computational puzzles. The first node to solve the puzzle receives a reward in the form of newly created or "minted" bitcoin \cite{narayanan2016bitcoin}. Referred to as "mining," this process could in theory be accomplished by any actor in the network. However, mining cryptocurrency on many blockchains is prohibitively expensive for individuals whose likelihood of finding the next block and securing the associated reward is vanishingly low \cite{beikverdi2015trend}. As a result, many miners join mining pools. Under the direction of a pool manager, mining pools work together to mine new blocks and distribute any rewards to the pool of workers \cite{narayanan2016bitcoin}. Over time, mining has shifted away from being performed by a network of distributed individuals, to being performed primarily by coordinated mining pools. Consequently, relatively few entities are responsible for the majority of mining power in both Bitcoin and Ethereum \cite{gencer2018decentralization}. Thus, even though a protocol may remain technically decentralized, economic forces outside the protocol nevertheless exert a centralizing force \cite{beikverdi2015trend}.  

The rules specified in a network's protocol can also be subject to outside influence. These rules are not static and self-contained; they are often updated over time based on the decisions of governing bodies and core developer teams or in response to regulatory action \cite{bodo2021decentralisation}. In other words, both the network topology and the network protocols are open systems subject to social, political, and economic forces that originate outside the network and can affect the degree and level of centralization within a system.

\section{Methods}
\label{sec:Methods}
\subsection{Participants}

In our research, we chose to focus on two kinds of blockchain and Web3 actors: 1) those contributing to blockchain platforms' technical or business development, hereafter referred to as "operators;" and 2) those investing in blockchain protocols through (venture capital) investment firms or as angel investors, hereafter referred to as "investors." We recruited participants initially by reaching out to contacts who have participated in the authors' research on similar topics in the past and by recruiting participants through industry conferences, Web3 events, and local meetups. From these seed contacts, we recruited additional participants via snowball sampling. See the Appendix for more details on recruitment. Rather than focusing on any single industry subsector or geography, we intentionally recruited participants working on diverse blockchain technologies to gain insight into themes that recur across the overall sector. In total, we conducted 30 to 90 minute interviews with 8 investors and 21 operators located in North America (\textit{n}=22), Europe (\textit{n}=5), South America (\textit{n}=1), and Africa (\textit{n}=1). Throughout the text, we indicate which participants' interviews in aggregate support our statements or correspond to specific quotations in brackets (e.g., [P1] would indicate a quote was from participant 1). We achieved theoretical saturation (i.e., the point at which we surfaced no new conceptual information) after these 29 interviews, and therefore ended recruitment at that time. We also conducted field observations at several industry events (DCentral 2022, Consensus 2022 and side events, the panel mentioned above on ESG and crypto hosted by VentureESG, the Princeton University DeCenter Inaugural Kickoff event, and Consensus 2023). Our analysis is concentrated on findings from our interviews with participants; however, we used our observations at events to ensure that our chosen analysis themes were also relevant in other contexts. \footnote{We have provided quotes associated with each theme that did not risk identifying participants \href{https://drive.google.com/file/d/1oOUJrOE0Kd3arBPZj2ItF5rKFOJLC4NC/view?usp=sharing}{here}}.  

Our data collection took place between May and December of 2022. During this period, the value of many blockchain-related assets sharply declined. For example, from the high in November of 2021 to June of 2022, the price of bitcoin fell from almost \$69K to below \$18K \cite{coinbaseBTC}. Furthermore, several high-profile blockchain projects collapsed. In May, TeraUSD--a stablecoin whose value would supposedly be stabilized by an algorithmic mechanism--collapsed, sending ripples throughout the sector  \cite{ostroff2022terra}. Subsequently, the centralized cryptocurrency lending service Celsius Network folded after making risky loans backed by very little collateral \cite{gladstone2022celsius}. These collapses foreshadowed the high-profile bankruptcy of the centralized exchange FTX and its associated hedge fund Alameda Research in late 2022. Sociotechnical scholars have pointed out that crises often make visible tensions and processes in technology communities that are opaque during times of stability \cite{defillipi2016invisible, walch2019deconstructing, dupont2017experiments, jabbar2019blockchain}. Thus, interviewing people during the beginning of the crypto market crash, or "crypto winter," offered a unique avenue for understanding how blockchain actors' values are (or more often are not) challenged by crisis. The Appendix details how our data collection lined up with several major industry collapses.

\subsection{Interview Protocol}
At the beginning of the interview session, we read participants a verbal consent script detailing our privacy protection practices and their rights as research participants. After participants verbally provided consent, we guided them through our semi-structured interview. Although the specifics of the interview varied somewhat based on participants' roles and experiences, in general, our interview questions focused on three core categories: 1) the goals of participants' organizations and their function within them; 2) participants' values and mission as related to the Web3 and blockchain industry; and 3) participants' ideas about investor-operator interactions. Participants were sent a gift card worth \$25 or the equivalent value in their country's currency. In cases where participants consented to having their interviews recorded, interviews were transcribed by a third-party transcription service. Our protocol was approved by the Princeton University institutional review board (IRB).

\subsection{Analysis}
We based our data coding method on abductive analysis \cite{tavory2014abductive, timmermans2012theory}, a qualitative data analysis method in which the analyst iterates between codes derived inductively from review of the data, as in grounded theory methods, \cite{charmaz2014constructing, strauss1990basics, glaser1978theoretical, glaser1968discovery}, and codes based on prior empirical and theoretical research that have importance to the research topic at hand. During initial coding, both authors reviewed the transcripts and independently applied descriptive codes, focusing on concepts that the authors identified based on prior research (e.g., decentralization, governance, privacy) as well as concepts that arose inductively from the interviews themselves (e.g., custodial versus non-custodial accounts, scams and Ponzi schemes). After discussing the resultant codes, the authors then proceeded to axial coding. During this phase, the authors identified higher-order categories that linked descriptive codes together and that had theoretical relevance. For example, the descriptive codes of "privacy," "custodial versus non-custodial accounts," and "ownership of data" all relate to the higher-order category of "self-sovereignty," which has been explored previously in analyses of blockchain technologies \cite{allen2016sovereignty, laatikainen2021towards, shrestha2019sov}. In the thematic phase of coding, we grouped axial codes into the four core themes: 1) promises: the current-day ills that Web3 and blockchain technologies seek to remedy (e.g., corruption and incompetence in Big Tech, traditional finance, and national governments) as well as the perceived societal benefits that can be achieved through these systems in the long-term; 2) challenges: the stumbling blocks that have or might hamper progress towards promises in the short-term; 3) perils: actual or possible harms that have arisen as the industry attempts to navigate stumbling blocks in the service of promises; 4) solutions: proposed approaches to mitigate perils and factors that might prevent their efficacy. The axial codes grouped within each of the four themes are depicted in Table \ref{tab:themes}. We achieved consensus on codes during each phase of coding through frequent, iterative discussions. \footnote{We did not calculate inter-rater reliability. Inter-rater reliability is methodologically unsuitable for interpretive qualitative research where codes arise from the collaboration between researchers and review of relevant literature rather than through techniques intended to surface statistically generalizable patterns \cite{mcdonald2019reliability}.}. Briefly, we note that participants disagreed in their definitions of core terms such as "Web3" and "decentralization" and in what they saw as the lines that divided different subsets of the broader sector. Throughout the results, we refer to "Web3 and blockchain" technologies to respect these distinctions. More information about definitions is available in the Appendix.

\bgroup
\def\arraystretch{1.5}
\begin{table}
  \caption{Analysis Themes}
  \label{tab:themes}
  \begin{tabular}{|p{0.12\textwidth}|p{.6\textwidth}|}
  \hline
    Themes & Axial codes\\
    \hline
    Promises & \begin{itemize}
        \item Self-sovereignty
        \item Participatory governance
        \item Financial access and inclusion
    \end{itemize} \\ \hline
    Challenges & \begin{itemize}
        \item Balancing organizational efficiency with stakeholder participation
        \item Expansion to mainstream markets
        \item Financing dynamics
    \end{itemize} \\ \hline
    Perils & \begin{itemize}
        \item Corruption and incompetence in governance and leadership
        \item Magnification of inequality
        \item Schemes and scams
    \end{itemize} \\ \hline
    Solutions & \begin{itemize}
        \item Consumer education
        \item Regulation
        \item Recommitment to the original blockchain ideals
    \end{itemize} \\ 
    \hline 
  \end{tabular}
\end{table}
\egroup

\section{Results}

\subsection{Promises}
Participants frequently discussed the benefits of blockchain and Web3 technologies, which included: \textbf{1) self-sovereignty; 2) participatory governance; and 3) financial access and inclusion}. 

Consistent with prior scholarship on self-sovereignty \cite{allen2016sovereignty, laatikainen2021towards, shrestha2019sov}, our participants thought of self-sovereignty as a multifaceted concept encompassing transportable identities [P3, P4, P5, P14, P18], interoperability [P6, P11], and privacy [P5, P6, P13, P16, P24, P25, P27, P29]. The most commonly discussed aspects of sovereignty were related to ownership and control of data and financial assets. Participants felt that Big Tech companies have gained too much power, especially through control of users' data. Facebook was a frequent target of excoriation [P1, P5, P6, P12, P19, P28, P29], with one participant describing it as an "evil empire [P12]," and another asserting that he hoped that "Mark Zuckerburg [would be] exiled to fucking Mars [P5]." Google was also frequently criticized as being a "Big Brother [P13]" that violates users' rights to privacy [P1, P6, P26, P28, P29], and as being a monopolistic juggernaut that "basically owns the internet [P28]." As a contrast to these practices, participants reasoned that decentralized technologies that placed ownership of data directly under the control of users would give users more agency over how it is (and is not) used. Unlike Big Tech companies that sell users' data to other platforms, service providers, and advertisers, several participants suggested that decentralized technologies could instead allow users to monetize their own data [P27, P23, P13]. 

Similar sentiments about direct ownership emerged around user control of financial assets. Participants castigated the traditional finance sector for being motivated by greed and characterized by widespread corruption. Several participants cited the bad behavior of the finance profession during the 2008 financial crisis as personal motivation to get involved with blockchain and Web3 technologies [P15, P20, P23]. Worse still, participants saw traditional finance companies as lacking accountability because their internal practices are a black box to consumers. Moreover, participants viewed the traditional finance sector as a "rent-seeking" intermediary whose primary motive was to extract financial gains from consumers unfairly. In contrast, participants saw decentralized technologies as liberating users from reliance on third parties like banks to intermediate transactions, eliminating the need for go-betweens who offer limited value and sometimes behave in untrustworthy ways. 

Yet participants took pains to point out that not all platforms that involve blockchain assets (e.g., cryptocurrencies or NFTs) are decentralized. Participants discussed this dichotomy in the distinctions between so-called "custodial" and "non-custodial" technologies. One participant explained: 

\begin{quote}
    "A custodial [model] would be when someone has protection of your assets and they're holding it in your custody, [but] they have access to it [...]. And the non-custodial is when you have [your currency] in a wallet that you have both the public address and the private keys to, and no one else has them except you [P27]."
\end{quote}

Many participants saw custodial models (i.e., intermediated models) as replicating many of the same problems that afflict traditional financial technologies because the user can only engage with their assets via a middleman. In contrast, decentralized, non-custodial protocols, sometimes referred to as "self-custody" models, were seen as better supporting user self-sovereignty because they allow parties to transact "trustlessly" insofar as transacting parties can "verify that everyone's work is correct [P4]" without need for a third-party intermediary.

In addition to ownership, participants felt that decentralized systems could offer users a second major benefit: participatory governance. Users have limited input over the platform's design and functionality within traditional technology platforms and finance companies. On the other hand, participants thought of blockchain and Web3 platforms as offering diverse platform stakeholders a mechanism to impact how platforms operate. To this point, several participants [P1, P18] who themselves were strong advocates for a collectivist perspective emphasized that blockchain and Web3 platforms could embody the ideal of participatory governance either by allowing platform stakeholders to directly vote for technical and policy changes or by electing stewards from amongst their community to represent platform stakeholders' interests. As a result, users, rather than internal management teams, could shape the trajectory of the platforms in ways that reflect their own values and priorities.

The third promise frequently discussed by our participants concerned financial access and inclusion {[P3, P4, P6, P7, P8, P10, P12, P14, P15, P17, P19, P20, P21, P23, P24, P26, P29]}. Although participants' perspectives on the role of policy and regulation were by no means a monolith, some participants viewed ineffective or incompetent governments, especially around the management of national economies, as one of their motivations for working on Web3 and blockchain technologies. Several participants noted that governments and central banks exert influence over national economies in ways that citizens disagree with or negatively affect citizens' financial well-being. In particular, participants raised concerns about high inflation [P1, P8, P12, P15, P16, P17, P18, P21] with some attributing inflation to poor government decision-making. Some emphasized that even if these effects were tolerable within Western contexts where economies are relatively stable, this is not necessarily the case in other geographic areas [P8, P10, P15]. From this perspective, because of the inadequacy of governments in effectively managing national economies, individuals do not have the freedom to make financial choices that are in their own interests. Thus, participants believed there is a need for a "worldwide financial system that's recognized in every country that isn't controlled by one government or entity that can determine who can use it and who can't [P12]."

Stemming from these motivations, participants viewed decentralized technologies as supporting financial access and inclusion in one of two ways. The first was access to stable capital, loans, and asset storage, which are typically afforded to citizens of developed counties, but not necessarily to citizens living in countries with developing economies [P1, P3, P4, P6, P7, P8, P10, P12, P14, P15, P16, P17, P18, P20, P21, P23, P24, P26]. On this point, several participants provided anecdotes about how blockchain technologies like cryptocurrencies supported precisely these types of people. For example, one participant cited an example of a friend who had spent time in a refugee camp. Because his friend owned Bitcoin, he was able to engage in commerce in ways that were not possible for other members of the same camp [P14]. The second way concerned giving users of all nationalities access to investment tools that would typically only be available to people in the financial sector or to the already very wealthy [P19, P20, P24, P26, P29]. According to some participants, the strategies that have allowed the wealthy to become even wealthier have not been available to everyday investors, a wrong that could be set right by technologies such DeFi, which (at least in theory) provides the same opportunities to everyone.

\subsection{Challenges}
\label{sec:Challenges}

Participants recognized that the path towards realizing the full promise of Web3 and blockchain technology is littered with potential roadblocks that could impede progress. \textbf{The three primary challenges discussed by our participants included: 1) managing the trade-off between organizational efficiency and participatory governance; 2) deterioration of the original blockchain vision when expanding to mainstream markets; and 3) challenges to ideals raised by platform financing dynamics.}  

The first challenge participants frequently raised concerned the tension between the need for organizational efficiency and the ideal of participatory platform governance. Operators of platforms that were still in their earliest stages, which constituted the majority of operators we interviewed, found this challenge particularly salient. For these platforms to establish product-market fit, they need to iterate quickly in response to market conditions and user feedback, which decentralized governance processes cannot easily facilitate. Thus, early-stage ventures often maintain some central leadership groups until they can transition to full community governance. As one participant explained:

\begin{quote}
"In Web3, you very often have this dual structure where there's a nonprofit protocol that maintains the technology and a for-profit foundation, [...] which works to steward the ecosystem [P1]."    
\end{quote}

While this dual structure can enhance quick decision-making, participants noted that organizations in this model are "not really decentralized [P9]." As a result, such organizations are still susceptible to the same foibles as traditional management and development teams that decentralized governance was intended to prevent. Consequently, many operators viewed the dual structure only as a transitional phase and aspired to eventually shift to full community governance, sometimes referred to as "exit to community."


When pressed on specifics of how platforms can eventually exit to community, operators rarely provided practical details and milestones for their own organizations. However, several pointed to the abstract roadmap laid out by a popular blog post from the prominent VC firm Andreessen Horowitz (a16z) entitled \textit{Progressive Decentralization} \cite{walden2020progressive}[P3, P5, P26, P29]. Under this model, platforms maintain central leadership teams early on to help establish product-market fit, then gradually involve the stakeholder community more once traction is achieved. Eventually, once an active user and technical contributor group is established, the central leadership teams can be dismantled and replaced with decentralized governance enacted via participation from the community of platform stakeholders. According to the post and our participants, this model has the added benefit of reducing the risk that any platform assets would subsequently be classified as an illegal security offering by the US Securities and Exchange Commission (SEC) since prior statements from regulators suggest that projects that are not run by a centralized group do not meet the criteria for being classified as a security \cite{hinman2018, walch2019deconstructing}.
 
Not all participants believed full community governance was achievable or even preferable, and some argued explicitly against it. One participant observed that humans are inherently hierarchical and largely ineffective when attempting to organize themselves otherwise [P9]. As evidence, she described a platform she worked with that had failed to update its technology due to the community's inability to reach a consensus on how to do so. Furthermore, some participants noted that members of the community may not always prioritize the long-term health of the platform, but instead their own self-interest [P20, P27].

\begin{quote}
    "The moment you mention a DAO, [developers] all scramble [because] it means [stakeholders] get to give me as much feedback as they want because they own a ton of tokens. They don't necessarily know what's good for the market. I find this unfortunate, but a lot of the voters basically vote for stuff that's going to make them more money [P20]." 
\end{quote}

In other words, conflicts can arise between an operator's techno-utopian ideal of collective platform governance and stakeholders, who may not themselves prioritize the utopian ideal unless it also serves their own financial goals.

The second challenge concerned the deterioration of the original blockchain vision when expanding to mainstream markets. Participants expressed concern that mainstream adoption of Web3 and blockchain technologies by consumers who were not animated by participants' own ideological values could "pervert" the vision for these technologies, especially if these audiences are primarily allured by the prospect of "fast bucks [and the] American dream [P12]." VCs and operators cast aspersions at sector speculators, sometimes referred to as "degens" \footnote{As one participant explained, "Degen is kind of a slagging off nickname that the community then adopted.  Degen [means] degenerative, only interested in money [P1]."} as well as the platforms that court them. Many participants thought that a benefit of the current market downturn in the Web3 and blockchain space would be to deter degens' and speculators' involvement in the sector altogether. 

How the sector can achieve mainstream adoption of (largely) financial technologies without attracting users with primarily financial motivations was not entirely addressed by our participants; however, some proposed that technologies that do not fully embody the Web3 and blockchain ideals could act as a tool for "onboarding" users with non-ideological motivations. Once onboarded, these users might see the value of technologies more aligned with the overarching vision. Several participants discussed centralized, custodial cryptocurrency exchanges as examples of such mechanisms. As another instance, one participant pointed to the success of Donald Trump's digital trading cards:

\begin{quote}
    "They comically onboarded a lot of normies, or people that don't know anything about crypto. And if you look on chain, a bunch of Trump trading card wallets that are holding these NFTs effectively don't have any gas \footnote{According to the Ethereum docs, "Gas refers to the unit that measures the amount of computational effort required to execute specific operations on the Ethereum network. Since each Ethereum transaction requires computational resources to execute, each transaction requires a fee."}. [...] which means they have to learn how [to add gas]. So he onboarded a bunch of people in the dumbest way, but he did it [P27]." 
\end{quote}

The third challenge our participants explored concerned how blockchain platforms finance their development, which can pit investors' financial interests against operators' higher-order ideals. For some operators, the presence of VC stakeholders who wield considerable power over the platform's assets and governance threaten user sovereignty and democratic governance. As one participant phrased it: "If the goal is collective ownership, it's a very different thing when 20\% of your company is owned by one [VC] player [P1]." 

Undergirding this concern is the relationship between VC firms and the limited partners (LPs) who invest in their funds. In other words, VCs primary fiduciary obligation is to their LPs, not their portfolio companies. For this reason, it was not uncommon for operators to provide overt criticisms of VCs as "extremely selfish [P15]," as "mercenaries [P15, P20]," who did not necessarily share the values of the blockchain or Web3 community [P5, P15, P26]. Several operators who oppose the VC model have opted to bootstrap their operations or finance growth from public token sales, which two referred to as "fair launches" because VCs cannot buy tokens at a discount in advance of the public token sale as is typically the case in VC blockchain investment deals. However, other operators felt that public sales are too risky from a regulatory perspective, since some regulators have signaled that all public token sales indicate that platforms are offering illegal securities \cite{higgins_sec_2018}. Instead, they see accepting VC financing as a necessary evil for scaling in the current regulatory environment [P5, P16].

Though many operators we spoke with expressed "nervousness around venture [P1]," two had enthusiastically pursued venture financing [P28, P29]. As reasons, both cited the desire to develop their technology and scale their ventures rapidly. One argued that without VC, his team might be outpaced by competitors with similar technology who did accept VC funds. He also noted that whereas founders who had previously founded companies with successful exits might have the luxury of bootstrapping operations, most do not. By accepting VC financing, he could spread the financial risk associated with launching a project. 

Both pro-VC and anti-VC operators emphasized that not all investors are equal. Some participants pointed out that inexperienced founders might be attracted to hedge funds as investors because they offer large investment sums with limited due diligence. Still, such funds offer deal terms that can be at odds with the platform's long-term health. For instance, Alameda Research, a hedge fund that collapsed in scandal alongside its associated centralized cryptocurrency exchange FTX, was highlighted as a bad actor due to the specific deal terms it offered that put founders' platforms at risk:

\begin{quote}
    "Every project that I spoke to about Alameda [was] always talking about how Alameda was selling their tokens. [...] If you see Alameda on anyone's website where they made an investment into that project, every single one of those project [tokens] are probably [worth] 10 times less than what they listed at [P28]." 
\end{quote}

Beyond investment firm type, founders explained several additional strategies to avoid becoming financially entangled with investors whose motives conflict with their higher-level vision. Some founders viewed an investor's hyper focus on tokens as a red flag [P2, P5], as it suggests that such investors "do not value human autonomy" and are just "trying to just get [their] bags [P5]." Others preferred to seek out VCs who had previously been operators themselves, as they were more likely to focus on the product's goals and the outcomes of the users, and were capable of providing valuable guidance [P16, P28]. 

For their part, VC investors did not believe that their obligations to their LPs posed a significant threat to realizing founders' visions. One investor noted that VCs are incentivized to support founders' visions since not doing so could harm their reputation and thus make promising founders less likely to work with them in the future [P4]. Additionally, VCs argued that the terms of their deals prevented them from selling their tokens or "dumping" on founders soon after they make their investment [P11, P22]. Therefore, they are bound by technical or legal constraints to support founders over a longer period of time. Moreover, they contended that any actions VC firms took to undermine the price of a platform's assets in the short or longer term would ultimately harm their positions and returns for their LPs. Consequently, VCs believed that their firms were not only personally aligned with the founders' visions, but also structurally incentivized to support them.

VCs, like operators, recognized that founder-investor relationships in the Web3 and blockchain space present unique challenges especially around verifying the legitimacy of founders' credentials and expertise. Early blockchain supporters were driven by a desire to protect user privacy \cite{swartz2018bitcoin, narayanan2013happened}, leading to the development of technologies that allow for pseudonymous transactions rather than requiring transacting parties to reveal their real-world identities. Following this cultural legacy, some founders seek funding without revealing their real-world identities to would-be investors. While two VCs we spoke with didn't view pseudonymity as a problem and even invested in pseudonymous artists [P13] and pseudonymous founders [P19], two others saw founder pseudonymity as a significant risk [P11, P22]. One investor focused on climate tech emphasized the importance of traditional credentials in that field. He argued that without knowing the founders' identity, it is difficult to assess their ability to execute their plans successfully. As a result, he chooses to work solely with founders who have disclosed their identity and who have traditional credentials in climate science, even if their expertise in Web3 and blockchain technology is limited. As he phrased it:


\begin{quote}
    "I think some of the basics around tokenomics or [smart contract development] are straightforward to learn for a smart, thoughtful person coming into the space. [...] So to me, I'd much rather pull in best-in-class people from climate and have them learn about crypto than the other way around [P11]."
\end{quote}

Paradoxically, the "negative filter" he applies to his process of assessing potential Web3 and blockchain companies to invest in excludes founders whose expertise and cultural identity is primarily connected to Web3 and blockchain.

Another VC indicated that founders who claim to have specific expertise in Web3 and blockchain often have a limited understanding of these technologies [P22]. He gave examples of instances where CEOs or even CTOs couldn't explain their rationale for choosing to build their platforms on one blockchain over another, which is a critical decision for any project in the space because it often constrains the capabilities of the platforms built on it. As a vetting strategy, he also chooses only to work with founders who disclose their real-world identities and can sufficiently answer detailed questions about blockchain technology and business practices. When pressed on why he though founders without relevant expertise would attempt to start a blockchain business, he said that he believed many aspiring founders viewed starting a Web3 or blockchain company as a get-rich-quick scheme rather than a serious technical and business pursuit. At the same time, several VCs we spoke with during our fieldwork reported that they faced tremendous pressure from their LPs to invest quickly in blockchain companies, especially those offering assets such as cryptocurrencies. Thus, VCs must balance demands from their LPs to capitalize on froth within the industry while avoiding investments in companies whose founders were likewise trying to opportunistically capitalize on hype.

 
\subsection{Perils}
\label{sec:Perils}
Participants recognized that while some of the challenges of the space could be managed, a failure to do so adequately could result in negative consequences for users, operators, investors, or the space as a whole. \textbf{The three perils discussed most often by our participants included: 1) corruption and incompetence in governance and leadership; 2) magnification of financial inequality; and 3) the prevalence of scams, Ponzi schemes, and fraudulent ventures}. 

First, participants emphasized that a major challenge in running Web3 and blockchain organizations involves establishing the correct initial governance structure, which typically maintains some form of centralization. A peril associated with this centralization is that, as with traditional technology organizations, corruption or incompetence within leadership can have disastrous consequences for platforms and their stakeholders. Our participants claimed that the fall of Celsius Network served as a clear example of this problem, as its unaccountable leadership caused financial harm to users and eventually to the platform itself [P14, P15, P22, P26]. Celsius did not employ a public, transparent blockchain and used a custodial account model similar to traditional banks, preventing users from adequately assessing their potential risks before the collapse and from withdrawing their funds once the collapse began.

Yet even when platforms use public, transparent blockchains and have some degree of decentralization in their governance policies, human actors within the platform can still accumulate power. One participant [P18] related an experience where he collaborated with the foundation of a platform that claimed to be on the verge of hosting elections for their board but ultimately never did "because it [would make] the people running the system more accountable [P18]." He also noted that a single actor on the foundation sometimes acted as "a blocker in the system" by delaying the deployment of funds approved through the platform's governance votes. As a result, the platform resembled "a regular corporate structure" in which central leadership figures can subvert the will of other platform stakeholders while still appearing to adhere to decentralized governance policies. As another participant put it: "saying that we're going to be decentralized for anything other than [trivial applications] really means we're going to have an invisible hierarchy, and we're not going to tell you how it works [P5]."

 The second peril of Web3 and blockchain is economic inequality, which is influenced by visible and invisible power dynamics, as well as skewed platform ownership and market dynamics. Two of our interview participants highlighted that despite Bitcoin's technical decentralization, Bitcoin mining heavily favors organizations with significant computational resources, making it difficult for ordinary individuals to secure financial returns from mining [P23, P27]. Another operator added that similar financial inequalities are present in Ethereum:

\begin{quote}
    "The economic distribution of Ethereum is highly centralized. The people who are in positions of wealth before are likely in greater positions of wealth now. There is a small margin of people [...] who've had an opportunity to get in early, but Alpha \footnote{Within the Web3 and blockchain space, the term "alpha" is used to describe early knowledge of a new technology or financial asset.} is an asymmetric totem pole--the closer you are to the top, the more access you have to an opportunity to grow. And the closer you are to the bottom, the more likely you are to be the exit liquidity of the top [P5]."
\end{quote}

Despite some Web3 enthusiasts seeing the technology as a potential long-term solution that democratizes access to financial tools and opportunities to accrue wealth, a significant gap exists between the current reality and the idealistic vision.

Two participants described "toxic positivity" as a problem in the Web3 and blockchain sector that magnifies the dynamics that result in financial inequality. "WAGMI," or "we're all going to make it," is a common saying amongst enthusiasts that implies anyone can achieve financial success by investing in cryptocurrencies and other blockchain assets [P5]. Critics who oppose this sentiment are dismissed as spreading "FUD," or "fear, uncertainty, and doubt" and are sometimes met with the retort "HFSP," meaning "have fun staying poor" \cite{kuhn2021HFSP}. However, one participant pointed to the evident logical fallacy of the WAGMI mentality: "statistically, we're not all going to make it. Someone needs to be my exit liquidity [P5]." Thus, toxic positivity in the space can deter would-be critics from pointing out when projects have clear flaws, which might prevent novice blockchain investors from exercising more skepticism about potential investments.

The third peril--scams, schemes, and frauds--constituted the most frequently discussed peril of Web3 and blockchain technologies [P1, P3, P4, P5, P12, P15, P16, P18, P22]. Several of our participants felt it was incumbent upon themselves to resist toxic positivity in the space by raising red flags about dubious projects, especially when such projects had hallmark characteristics of being scams or Ponzi schemes. However, what participants considered scams or a frauds varied across participants. For instance, many participants labeled the collapsed project Terra as a scam, fraud, or Ponzi scheme [P1, P3, P5, P12, P18, P28]. As one participant phrased it:

\begin{quote}
    "I saw that shit from a mile away. It's an under-collateralized algorithmic stablecoin, my ass. No. It was like Magic Internet Money [like other well-known crypto schemes]. You know it's a Ponzi [P5]."
\end{quote}

While many participants viewed Do Kwon, the project's founder as "a total loose cannon [P1]" and "a criminal [P5]," others felt that he had not intentionally swindled retail investors out of their funds but rather, merely failed to create a sufficiently robust system [P4, P15, P16] \footnote{Importantly, all of our interviews were conducted before the SEC filed securities fraud charges against Terra and its founder Do Kwon in February of 2023. Our participants may have revised their opinions about Terra as new allegations of criminal conduct have come to light}. As one VC explained:
\begin{quote}
"I think some people have called what happened with Luna and UST a Ponzi or a fraud or a scam, and I actually disagree with that sentiment. Let me be very clear about parsing it through, [...], a Ponzi in my definition is a very clear fraud. It is me taking your money and giving it to someone else under the guise of interest or a return. That is quite literal fraud. Creating a new technology is not fraud. It could work. It could not work [P4]."
\end{quote}

Along similar lines, another participant who described Do Kwon as "a bit of an ass" but who nevertheless "inevitably inspired a generation of builders," argued that Terra was not a fraud, but a system that was toppled by "a long tail risk of their economic system," which revealed that "that the system was good, but not good enough [P16]."



Although these participants pointed to Terra as having unsustainable business practices, poor governance, or flawed technology, ultimately, they did not see it as a deliberate, intentional scam or Ponzi scheme. Some participants in our sample even conceded to themselves losing significant sums of money in the collapse of Terra. During our fieldwork, we heard similar stories from other industry insiders who, encouraged by the backing of respected investors \cite{miller2022terra}, also lost large investments in Terra. How everyday investors can be expected to determine which projects are legitimate and which are not is unclear, given that even industry insiders cannot do so consistently. As a result, the benefits of industry insiders calling out the "scammers and bullshitters [P22]" is inherently limited by the heterogeneity in what they consider to be scams and bullshit in the first place. 

\subsection{Solutions}
Participants also reflected on potential solutions to problems facing users and the blockchain and Web3 sector as a whole. \textbf{The three most commonly discussed solutions included: 1) education; 2) regulation; and 3) recommitment to the original blockchain ethos.}

Our participants frequently emphasized the importance of retail investors taking responsibility for self-education about the blockchain and Web3 projects they engage in [P5, P12, P14, P15, P21, P25, P26, P28, P29]. They expressed this sense of responsibility through the sentiment of "do your own research" (i.e., "DYOR"). However, the definition of DYOR was debated among participants, with one remarking that the lack of a standard definition was problematic [P27]. Participants described negative experiences as a vector through which they themselves or other sector consumers learn to DYOR. One participant shared his experience of falling victim to an NFT "rug pull" scheme, which motivated him to learn how to code in Solidity, the programming language used to develop smart contracts on the Ethereum blockchain [P12], to better evaluate his investment risks. Other participants also shared stories of getting scammed or losing money on blockchain projects, which was seen as a potential catalyst for becoming better educated about the technology [P5, P14, P15, P20, P27]. Thus, some viewed losing money on fraudulent or faulty projects as a way to learn about blockchain and embrace the maxim of DYOR.

Participants also suggested academic institutions and online resources could be a means for gaining knowledge [P15, P22, P25, P26, P28]. However, one participant raised concerns about the legitimacy of ostensibly educational content in the blockchain space [P22]. He drew our attention to the website of Blockgeeks \footnote{https://blockgeeks.com}, a blockchain education platform that prominently displayed an endorsement from Vitalik Buterin, the co-founder of Ethereum. The participant pointed out that Buterin's father owns and manages Blockgeeks, indicating a potential conflict of interest. As a result, this participant argued that "everything he teaches is automatically not impartial [P22]." Thus, while many participants considered education a critical component of user sovereignty, they were less clear on how impartial consumer education could be achieved within an environment marked by conflict of interest.

Participants discussed regulation as a second potential solution to address issues in the Web3 and blockchain space but disagreed on the role regulation should play in the sector. Some participants strongly opposed government regulation [P12, P15, P25]. In contrast, others believed that it could be beneficial in certain areas, such as consumer protection efforts that flag when projects advertise dubious yield "opportunities." Others expressed a desire for regulation to clarify what constitutes a security for digital assets and to better protect project developers [P16, P20]. Yet even within those areas, our participants expressed concern that regulators may not fully understand the technology well enough to draft thoughtful legislation and could end up favoring larger, established players who have more resources to exploit regulatory loopholes and suppress innovation [P5, P6, P15, P16, P19]. In sum, while in theory, participants were not entirely opposed to regulatory attempts to correct some of the sector's flaws, they offered few, if any, regulatory solutions that could be enacted in practice.

Participants also discussed the idea of a stronger commitment to the original principles of decentralization, self-sovereignty, and transparency that guided early blockchain advocates as another potential solution to many of the problems arising in the sector today. Some participants believed that doubling down on these principles could prevent the failures that contributed to the recent market turmoil. As examples, some participants pointed to what they saw as the successes of DeFi protocols compared to the failures of centralized counterparts such as Celsius and FTX. As one participant explained:

\begin{quote}
    "[DeFi platforms] all held their grounds really well [...] The total value locked up in those smart contracts at any given time and any movement in and out was visible to everybody. And I think that proved that governance managed by a set number of servers that are distributed actually really works [P15]."
\end{quote}

Participants saw the transparency of technologies with public blockchains as a potential deterrent to bad behavior or, at a minimum, as a means for other users to have more information about where the activity of other users might compromise their own positions. That is, it is easier to DYOR when information about transactions is openly available on the blockchain ledger.

\section{Discussion}

This paper explores the overarching vision of operators and investors for the future that could be supported Web3 and blockchain technologies. Participants believed that through decentralization, blockchain and Web3 technologies have the potential to improve user sovereignty, distribute governance power, and support more inclusive financial systems. However, they recognized that achieving these positive outcomes could be difficult. As challenges, they pointed to the difficulty of adopting technical and organizational decentralization early in platforms' development. They also highlighted that financing dynamics can introduce conflicts of interest between investors and operators that could impede progress towards ideological aims. Lastly, they expressed concern that as blockchain and Web3 technologies attract mainstream users, these users' financial motives may challenge the sector's ability to make progress towards more revolutionary goals. Participants emphasized that failure to address challenges effectively could further perpetuate existing negative outcomes in the industry including scams, fraudulent schemes, loss of funds for retail users, and the widening of financial inequality. Although participants discussed potential solutions such as regulation, consumer education, or a stronger commitment to the original ideals of the blockchain space, they expressed doubts about the efficacy of these approaches.

A unifying thread that connected all of the themes of our analysis is that participants did not see existing Web3 and blockchain technologies as achieving their utopian ideals and generally did not specify the practical path to doing so. On the other hand, blockchain and Web3 technologies have already caused and likely will continue to cause tangible harms. For example, Chainalysis reported that in 2022 \$3.8B were stolen from blockchain protocols \cite{chainalysiscrime}, and the popular anti-blockchain website "Web3 is Going Great" reported in July 2023 that "\$67,558,063,942 has been lost to hacks, scams, fraud, and other disasters in since January 1, 2021 \cite{whiteweb3goinggreat}." Reporting on the collapse of Terra indicates that retail investors suffered massive financial losses \cite{ostroff2022terra}, with multiple reports surfacing of suicides linked to the platform's meltdown \cite{morriscryptocrooks}. 

While some participants expressed empathy towards retail investors who experienced financial losses from collapsed projects, at least as often participants were dismissive of these events and chided retail investors for being under-informed. In other cases, participants minimized the impacts of scams, hacks, and market volatility in the space by arguing that these events receive disproportionate attention from the media and are not unique to the Web3 and blockchain sector [P1, P11, P25], ignoring the distinction that victims often have a mechanism for recourse in other sectors. 

Despite generally feeling that retail investors should accept responsibility for the consequences of their choices good and bad (i.e., "DYOR"), participants recognized that the sector-wide prevalence of scams, high-profile project failures, and unethical business practices does represent a threat to the space's long-term flourishing. At the same time, participants rarely embraced solutions such as regulation that would have sector-wide impact, in part because many felt that regulation runs counter to their beliefs in individual sovereignty and free markets. Instead, they embraced solutions that were more consistent with their own value systems such as consumer education, which shifts the burden of responsibility onto end users. Given that many industry-insiders we spoke to during our interviews and at industry events admitted to themselves falling for scams or too-good-to-be true investment schemes, it is unclear how effective consumer education will be in mitigating these issues. Summarizing a similar clash between idealized notions of individualism and the reality of harms during the 2017 ICO bubble, Swartz concludes, "Crypto promises freedom, including the freedom to scam and be scammed \cite{swartz2022theorizing}."

Many participants believed that democratic participation in governance is essential for a new era of technology in which the needs and desires of stakeholders are prioritized over those of corporate actors. However, some participants questioned whether novel organizational governance models would actually be in the best interests of all stakeholders [P7, P9, P20, P29]. Some went so far as to argue that because the structure of traditional corporate governance has been improved over time, it is already well-suited to serving capitalist systems [P7, P29]. Several participants also cited instances where decentralized governance compromised progress within the platform [P9] or allowed malicious actors to manipulate governance for personal gain [P27]. Thus, while participants generally supported the pursuit of democratized platform governance, they acknowledged that these forms of governance are still experimental and have not yet achieved the desired results. Whether such governance mechanisms will eventually mature to support the utopian ideal remains an open question.

Some participants saw transparency as a bulwark against would-be malicious actors since such behavior could be readily detected by anyone inspecting the ledger of transactions. In support of this view, several participants pointed to the abuses of centralized platforms like Celsius Network and FTX whose opaqueness prevented stakeholders from unveiling the poor risk management and unscrupulous business practices that compromised users' investments [P14, P15, P22, P26, P27, P28, P29]. On the other hand, the failure of Terra brings a sobering illustration of the limitations of transparency. As one participant noted, "this was transparent. We knew where the [investment protocol] returns were coming from. They were coming from [Terra's foundation]. It was all known [P4]." \footnote{Again, participants who did not see Terra as criminal or fraudulent at the time of their interviews may have revised their positions in light of new evidence and charges.}. Yet access to the ledger of transactions on the Terra blockchain as well as critiques pointing out the flaws in systems like Terra \cite{clements2021built, byrne2017stable} did not dissuade many retail users from attempting to capitalize on promised returns, and ultimately suffering massive financial losses. Moreover, transparency in the Terra blockchain is not equivalent to transparency into the technical and business practices of key figures within Terra. Since Terra collapsed, more information about unethical and potentially criminal behavior on the parts of these figures has come to light \cite{morriscryptocrooks5}. Without transparency into both blockchain transactions and well as internal practices, even the most diligent consumers cannot make fully informed choices.

As with consumer education, calls for transparency shift the burden of enforcing accountability to users of the system. Summarizing the limitations of transparency, Ananny and Crawford \cite{ananny2018seeing} write "If transparency has no meaningful effects, then the idea of transparency can lose its purpose [...] Transparency can reveal corruption and power asymmetries in ways intended to shame those responsible and compel them to action, but this assumes that those being shamed are vulnerable to public exposure." That is, for transparency to work in the service of accountability within blockchain systems, users must: 1) be able to make sense of blockchain transactions or other public source code; and 2) have a mechanism to hold bad actors accountable if problematic behavior is detected \cite{ananny2018seeing}. Human-computer interaction researchers have only recently begun considering human factors for blockchain and Web3 technologies \cite{elsden2018making}, but evidence suggests that everyday users often lack understanding of how core properties of blockchain technologies work in ways that leave them vulnerable to financial loss \cite{gao2016two, mai2020user, voskobojnikov2020surviving, voskobojnikov2021u}. Moreover, given the current level of regulatory ambiguity in the space, the prospect of legal consequences may or may not deter bad actors. As a result, the impact of transparency on actor behavior is murky at best, especially in the absence of policing by regulators, which many of our participants opposed. 

Similar concerns arise around system designs that support a high degree of user sovereignty, especially non-custodial technologies in which users manage their own private keys. Over the last few decades, the privacy and security community has begun to recognize that the best security practices incorporate an understanding of human users' needs, motivations, and cognitive limitations \cite{ackerman2005privacy, cranor2005security, iachello2007end}. Without a deep awareness of human factors, system developers can design systems that actually hinder rather than promote privacy and security. For example, even if systems adopt complex password schemes to support the highest possible security, these good intentions can be subverted if users seek to manage the complexity of higher demands in ways that compromise security \cite{lee2022password}. During Consensus 2023, the first author of this work attended a panel about self-custody in which the panelists asked the audience to raise their hand if they had lost crypto assets due to mismanagement of private keys in non-custodial systems. Nearly half the audience raised their hands, dovetailing with findings from prior literature examining the usability challenges of non-custodial technologies \cite{frohlich2020don}. As with democratic governance and system transparency, it bears underscoring that systems that could theoretically support self-sovereignty may not actually do so when they are deployed in the realm of human actors who are characterized by a variety of cognitive and behavioral foibles. 

Many of the challenges and perils identified by our participants pertained to how financial incentives drive behave in ways that undermine the intended social mission of technology platforms. Prior scholarship on blockchain cultures demonstrates that a significant faction of blockchain advocates are animated by a belief in the fundamental superiority of free markets \cite{swartz2018bitcoin, narayanan2013happened, cousins2019value}. Some of our participants likewise believed that cryptocurrencies, DeFi, and other financial blockchain technologies should not be regulated and that regulation would hinder competition, innovation, and efficiency within markets. Yet the reality of these systems suggests a different equilibrium. Blockchain critic Herian argues that the outcome of unregulated markets is not fierce competition, but further entrenchment of power \cite{herian2018blockchain}. In a related analysis, Crandall \cite{crandall2019blockchains} argues that the dynamics of the growing crypto industry in Puerto Rico are akin to the 2008 financial crisis. As in 2008, powerful crypto players seek to leverage institutions for their own gains while promoting the rhetoric of the superiority of free markets. In this way, blockchain and Web3 efforts are more aligned with neoliberal than libertarian agendas because, like prior neoliberal attempts to entrench free-market ideologies via control of institutional actors \cite{mirowski2014never}, they do not disrupt power and wealth structures, but further ossify them. Given the ever-expanding financial inequality in blockchain systems, evidence for Herian's and Crandall's positions is compelling whereas evidence that these systems promote financial inclusion is limited. 

Our participants saw rampant speculation as one of the barriers to progress in the industry. This criticism, however, ignores several important realities. The allure of potential profits is one of the core motivations for users to adopt blockchain technologies, especially cryptocurrencies and other blockchain digital assets \cite{frohlich2020don, mattke2021bitcoin, cousins2019value, fan2023altruistic, hoyng2023bitcoin}. Arguing that actors with financial motives risk undermining the higher purpose of Web3 and blockchain technologies is akin to proclaiming that customers who are hungry are an affront to restaurateurs. Furthermore, incentive engineering, referred to by two participants as "mechanism design" [P7, P11] is why many blockchain advocates believed these systems would work to being with \cite{buterin2017introduction, crandall2019blockchains}. If it is in an actor’s best interest to behave "honestly"--that is, in accordance with the protocol--they are likely to behave in ways that support what the system designers intended. Yet designing systems whose incentive structures are complex enough to account for every possible way actors might attempt to undermine the designer's intention is a tall order as the failure of Terra illustrates quite clearly. Moreover, decades of research in the field of behavioral economics has long documented that humans do not always behave in economically rational ways \cite{kahneman1979prospect} both because they do not have access to the infinite cognitive resources to determine the optimal action \cite{simon1957} and because they are compelled by both economic and non-economic motives \cite{fehr2007social}. Systems whose proper functioning relies on the assumption that human behavior is rational are likely to face significant challenges.

Given the disconnect between the techno-utopian world our interviewees envision and the messy reality of today's Web3 and blockchain technologies, why does this vision persist in the minds of investors, operators, and many users? The technology industry has a long history of visionaries attempting to produce technologies that do not merely solve a simple problem, but engender social change. In the service of this aim, technologists often develop technologies that make some progress towards their goal while also creating other problems. For example, while social media is seen as exacerbating mental illness amongst teenage girls \cite{twenge2022specification}, engendering political radicalization \cite{hassan2018exposure}, and propagating misinformation \cite{lee2023social}, it also played instrumental role in social movements such as the Arab Spring, Occupy Wall Street, and Black Lives Matter \cite{skinner2011social, arafa2016facebook, mundt2018scaling}. In the same vein, our participants did not see their failure to achieve their loftiest ideals as a condemnation of their mission, and felt that continuing to push towards it was worthwhile despite the problems that abound. As one participant framed it, "there is a total ideal of what a utopian situations is like. But we are still helping out with the fundamental, broader changes of decentralization in access and so on. [...] To say the system failed because it didn't hit the utopian ideal, I think is bit harsh on what is really practically possible [P19]." In other words, participants felt that despite the problems raised by an incomplete realization of the Web3 and blockchain vision, progress towards this vision, however fraught with challenges and perils, is still a worthy goal.

\section{Limitations}
Our study has several important limitations. First, because we interviewed a limited number of participants recruited from specific events or via snowball sampling, the opinions and experiences of our participants may not generalize to all Web3 and blockchain enthusiasts. We note, however, that recruiting participants and conducting ethnographic fieldwork at public or semi-public industry events is a recognized mechanism of "studying up," that is, studying research subjects who have access to wealth and institutional power that are often unwilling to be observed in their day-to-day contexts \cite{souleles2018study}. Because we intentionally sampled participants working on or investing in a broad array of technologies, we were not able to meaningfully identify differences that might emerge between Web3 and blockchain subcultures (e.g., between DeFi communities and DAO communities). Future research could build on the foundation we have created in this work to to create a generalizable account of the opinions and beliefs of a larger group of actors through either quantitative survey analyses or larger-scale qualitative investigations. 

\section{Conclusion}
In this work, we have illuminated what Web3 and blockchain operators and investors see as the higher aim for their technologies in the future, what they see as problems and harms wrought by present-day attempts to make progress towards this future, and what they consider to be viable solutions. Our analysis can serve as a road map for the CSCW community for understanding how the designs of blockchain and Web3 systems fall out of a coherent set of beliefs centered around individual liberties and participatory governance. Our analysis can also provide insight into how this community is likely to think about future iterations on their technologies and their own responsibility (or lack thereof) in protecting their users from harms such as catastrophic financial loss. Furthermore, organizational sociologists have posited that when organizations are subject to regulations that run counter to their normative beliefs, they are less likely to internalize meaningful changes and more likely to comply only ceremonially \cite{oliver1991strategic, scott2013institutions}. Thus, our analysis may help regulators in understanding how to frame potential regulations to Web3 and blockchain companies and seek their participation in drafting regulations collaboratively.

\section{Acknowledgments}
We thank Sarah Scheffle, Andr\'es Monroy-Hern\'andez, and Klaudia Jazwinska for helpful feedback on our manuscript. We gratefully acknowledge financial support from the Schmidt DataX Fund at Princeton University made possible through a major gift from the Schmidt Futures Foundation.

\bibliographystyle{ACM-Reference-Format}
\bibliography{bibliography}

\input{appendix}

\end{document}

%% file: appendix.tex
\label{sec:Appendix}

\section{Recruitment}
Details about participant roles and geographic location as well as the method through which they were recruited and the date they participated are available in the table below. We note that the Price of TerraUSD fell to near 0 cents on May 12, 2023 \cite{sandorTerra2022}. Celsius Network froze transactions on June 12, 2022 and filed for Chapter 11 bankruptcy on July 13, 2022 \cite{NapolitanoCelsius2022}. On November 11, 2022 FTX filed for Chapter 11 bankruptcy in the United States \cite{DeFTX2022}.  


\begin{longtable}{|>{\raggedright}p{0.15\linewidth}|>{\raggedright}p{0.1\linewidth}|>{\raggedright}p{0.15\linewidth}|>{\raggedright\arraybackslash}p{0.3\linewidth}|>{\raggedright\arraybackslash}p{0.1\linewidth}|}
\hline
\rowcolor[HTML]{C0C0C0} 
{\color[HTML]{000000} \textbf{Participant}} &
  {\color[HTML]{000000} \textbf{Role}} &
  {\color[HTML]{000000} \textbf{Geography}} &
  {\color[HTML]{000000} \textbf{Recruitment Method}} &
  {\color[HTML]{000000} \textbf{Date}}\\ \hline
\endfirsthead
\endhead
P1                                                                 & {Investor}      & UK/EU              & {Research interviewee for earlier project of second author}     & {05-23-2022} \\ \hline

P2                                                                 & {Operator}      & UK                 & {Research interviewee for earlier project of second author}     & {05-23-2022} \\ \hline

P3                                                                 & {Founder}       & US                 & {Research interviewee for earlier project of second author}     & {05-23-2022} \\ \hline

P4                                                                 & {Investor}      & US                 & {Research interviewee for earlier project of second author}     & {05-23-2022} \\ \hline

P5                                                                 & {Founder}       & US                 & {Art Basel 2022}                                                & {05-25-2022} \\ \hline

P6                                                                 & {Operator}      & DE                 & {Personal contact of second author}                             & {05-25-2022} \\ \hline

P7                                                                 & {Partner}       & US/DE/dev world    & {Contact from non-profit VC ethics initiative of second author} & {05-31-2022} \\ \hline

P8                                                                 & {Investor}      & Africa             & {Research interviewee for earlier project of second author}     & {06-01-2022} \\ \hline

P9                                                                 & {Founder}       & US                 & {Meetup}                                                        & {06-24-2022} \\ \hline

P10                                                                & {Investor}      & US                 & {Contact of P7}                                                 & {06-28-2022} \\ \hline

P11                                                                & {Investor}      & US                 & {Contact of P7}                                                 & {06-29-2022} \\ \hline

P12                                                                & {Employee}      & Canada             & {Consensus 2022}                                                & {06-28-2022} \\ \hline

P13                                                                & {Investor}      & US                 & {Contact from non-profit VC ethics initiative of second author} & {06-29-2022} \\ \hline

P14                                                                & {Investor}      & US                 & {Contact from non-profit VC ethics initiative of second author} & {07-19-2022} \\ \hline

P15                                                                & {Founder}       & US                 & {DCentral 2022}                                                 & {07-20-2022} \\ \hline

P16                                                                & {Founder}       & US                 & {DCentral 2022}                                                 & {07-20-2022} \\ \hline

P17                                                                & {Operator}      & US                 & {DCentral 2022}                                                 & {07-22-2022} \\ \hline

P18                                                                & {Operator}      & Canada             & {Consensus 2022}                                                & {07-22-2022} \\ \hline

P19                                                                & {Investor}      & UK                 & {Contact from non-profit VC ethics initiative of second author} & {07-25-2022} \\ \hline

P20                                                                & {Founder}       & Canada             & {Consensus 2022}                                                & {07-26-2022} \\ \hline

P21                                                                & {Founder}       & Columbia           & {Consensus 2022}                                                & {07-27-2022} \\ \hline

P22                                                                & {Partner}       & UK                 & {DCentral 2022}                                                 & {07-27-2022} \\ \hline

P23                                                                & {Operator}      & US                 & {Consensus 2022}                                                & {08-09-2022} \\ \hline

P24                                                                & {Operator}      & US                 & {Consensus 2022}                                                & {08-09-2022} \\ \hline

P25                                                                & {Founder}       & US                 & {Contact of P26}                                                & {09-02-2022} \\ \hline

P26                                                                & {Founder}       & US                 & {Consensus 2022}                                                & {09-02-2022} \\ \hline

P27                                                                & {Founder}       & US                 & {Contact of P26}                                                & {12-20-2022} \\ \hline

P28                                                                & {Founder}       & US                 & {Contact of P27}                                                & {12-20-2022} \\ \hline

P29                                                                & {Founder}       & US                 & {Contact of P27}                                                & {12-20-2022} \\ \hline
\end{longtable}

\section{Methodological Justifications}
Qualitative methods utilized in social and cultural analyses have been a subject of contentious debate regarding their validity and usefulness, especially in the past two decades \cite{wuthnow2011taking}. One particular aspect that has drawn criticism is the use of semi-structured interviews, which are commonly employed in sociocultural studies. Some critics even go to the extent of claiming that these interviews offer no value at all \cite{jerolmack2014talk}. These critiques often originate from scholars associated with quantitative cognitive sciences and are based on the assumption that consciousness can be divided into a readily accessible and expressed "discursive, surface level," as well as a more significant yet challenging to access "deeper, visceral level" that influences behavior \cite{pugh2013good}(p.45).

From this standpoint, quantitative methods like behavioral or neural experiments and surveys that tap into the automatic and unconscious aspects of visceral consciousness are considered more valid for evaluating cultural and social knowledge, beliefs, and actions. Proponents of this position point to various studies that demonstrate a discrepancy between what research participants say and what they actually do \cite{vaisey2009motivation}.

Setting aside the fact that quantitative behavioral methods are not immune from a variety of idiosyncrasies that impact results \cite{nicholls2006satisfaction, stangor2014research, podsakoff2003common}, defenders of semi-structured interview methods point out that such critiques also miss the point of using such methods to begin with. Advocates of qualitative interviewing approaches point out that interviews can offer several types of knowledge, which sometimes go beyond that which can be ascertained through quantitative techniques. For example, according to Pugh \cite{pugh2013good}  interviews can surface four types of information: 1) “the honorable,” meaning information on how interviewees present themselves so as to be consistent with notions of what it means to be noble or admirable; 2) “the schematic,” meaning information about how interviewees convey their viewpoint such as through jokes or plays on words; 3) “the visceral,” meaning the underlying emotions and beliefs that influence interviewees behavior; and 4) “meta-feelings,” meaning “how we feel about how we feel.” Pugh argues that these types of information need not be consistent with each other, and instead argues that contradictions can be important for highlighting the emotional landscape that contributes to how interviews engage in the process of meaning-making, highlighting interviewees’ uncertainties and anxieties as well as the problems they face within their cultural environments. Particularly when paired with interpretive analysis techniques such as the ones we used in the current study, interview studies can go beyond the content of interviewees’ words so as to ascertain their deeper meaning.  

Other qualitative theorists have pointed out that whereas quantitative methods such as surveys and experiments are useful for evaluating evidence for a researcher’s specific question (i.e., deductive approaches), they are less well-suited to exploratory research. Inductive \cite{strauss1990basics, glaser1978theoretical, charmaz2014constructing} or abductive \cite{timmermans2012theory, tavory2014abductive} qualitative methods can allow insights to surface through analysis of the data that otherwise would not have arisen if the data collection and analysis methodology had been designed specifically to answer one or a few specific questions. In other words, quantitative, deductive techniques do not provide any answers to questions that the researcher did not explicitly ask. Given that blockchain and Web3 technologies and their associated communities are still in their infancy and relatively little research has been conducted on their social, cultural, and organizational dynamics, the use of exploratory and interpretive approaches to data collection and their analysis is appropriate. 

Other critiques of qualitative methods concern the extent to which such methods should comply with quantitative fields’ move towards open/transparent and reproducible methods \cite{pratt2020editorial, murphy2021ethnography, rhoads2020whales, kapiszewski2021transparency}. With regards to transparency, methods to protect participants’ anonymity pose a particular challenge. Participants in qualitative interview studies often consent to participate under the condition that their anonymity will be maintained, as was the case in our study. This often involves replacing information that might individually identify participants with more general descriptors or obscuring such information in some other way. However, this can raise questions as to the veracity of information conveyed within publications using such strategies, especially if the methods used to obscure identifying information substantively alters the meaning or interpretation of the data \cite{rhoads2020whales}. One solution to this problem would be to only interview participants who are willing to have their identities and their interviews transparently disclosed. Yet this strategy severely constrains what qualitative researchers can address in their research, especially if their research questions involve an exploration of topics that might compromise participants’ safety, security, or well-being if discussed openly. Here, we explicitly sought to address topics that could uncover social, cultural, and economic tensions as experienced and expressed by blockchain and Web3 operators and investors; thus, we chose to interview our participants under the condition of anonymity so that they could discuss these difficult topics without fear of rebuke or retribution from investors, collaborators, or their broader communities. 

Even though we chose not to disclose participants’ identities here, we recognize that transparency into the research process is important for enabling scrutiny of our methods and results and potentially for enabling secondary analysis by other researchers. To facilitate this, we provide extensive supporting quotes for our themes where such quotes did not risk identifying the interviewees \footnote{Quotes can be accessed \href{https://docs.google.com/spreadsheets/d/18zpVsnsC-l7nqJKG1PyOGdh4OBYDSjS4yvz720t-cT0/edit?usp=sharing}{here}}. We note, however, that this effort to increase transparency is not intended to support reproducibility, a value that has become increasingly important in quantitative research. Interpretive analyses are by-design and by necessity dependent on the perspective of the researchers performing them. Reflexivity, or “a focus on who I am, who I have been, who I think I am, and how I feel affect[s] the data collection and analysis \cite{pillow2003confession}(p.176),” is an important practice in general in the social sciences for contextualizing how researchers’ own perspectives might affect their research, but especially for qualitative studies. Without devolving into unnecessary navel gazing on positionality, we note that the first author has academic training in quantitative psychology and neuroscience. She has worked as a practitioner of data science in the technology industry, and in her current role has published both quantitative and qualitative research on machine learning and the practice of machine learning within applied settings. The second author has an academic background in economics and anthropology. He has professional experience in management consulting as well as in venture capital investing and has published ethnographic studies of both homelessness and venture capital. Given the same data, researchers with different backgrounds would likely derive different conclusions or find different aspects of our interviews meaningful. We hope that interested readers will do exactly that.


\section{Defining Core Terms}
Participants had differing definitions of industry terms, particularly regarding Web3 and decentralization. Ownership was frequently mentioned as a defining element of Web3 by several participants [P2, P13, P14, P23, P26, P27, P28, P29], while others emphasized coordination of stakeholders [P11, P26] or promotion of the common good [P1, P18]. There was no consensus on the centrality of blockchains within Web3 technologies, with some participants explicitly defining Web3 based on blockchains [P11, P23, P29] and others providing conceptualizations based on value systems [P2, P13, P14, P23, P26, P27, P28, P29]. One notion of Web3 compared it to Web1 and Web2 where Web1 enabled "read" only, Web2 enabled "read" and "write" and Web3 will enable read, write, and "own" [P27]. 

Participants also showed no consensus regarding which specific technologies should and should not be considered a part of Web3. Participants who argued for values-based definitions of Web3 pointed to the InterPlanetary File System (IPFS) [P27], Torrentz [P28], and Telegram [P28] as examples of Web3 technologies without blockchains. Even within technologies that leverage blockchains, there was disagreement on what should "count" as Web3. For instance, one participant explained that Web3 and cryptocurrencies are often incorrectly conflated, including how we, the authors of this work, used these terms in our interview with him. To provide an example of his delineation, he argued that while decentralized finance (DeFi) technologies can be considered cryptocurrency technologies, they should not be considered a part of Web3 [P4]. On the other hand, two founders working on DeFi platforms specifically designated their platforms as following a "Web3 ethos" [P20, P26], in that they redistribute ownership, agency, and financial benefits back to the community of users. 

There was also a variety of definitions regarding decentralization. While some participants gave simple definitions of decentralization related to the removal of central authorities or intermediaries and the distribution of decision-making power across the network of nodes [P17, P26, P28], others argued that such definitions were too simplistic. These participants raised concerns about formal and informal organizational processes, distribution of financial benefits, and power dynamics. For example, two participants [P23, P27] cited Bitcoin to illustrate the nuance of defining decentralization. Both pointed out that while Bitcoin uses an architecture that is designed to be decentralized, the computational demands of mining Bitcoin has contributed to a consolidation of mining power within a very small number of entities who wield disproportionate power within the network and receive the associated financial returns. In other words, "from a technical perspective, is [Bitcoin] decentralized? Yes. From a behavioral perspective, is it decentralized? I don't think so [P23]." 


%% file: main.bbl

\begin{thebibliography}{113}


\ifx \showCODEN    \undefined \def \showCODEN     #1{\unskip}     \fi
\ifx \showDOI      \undefined \def \showDOI       #1{#1}\fi
\ifx \showISBNx    \undefined \def \showISBNx     #1{\unskip}     \fi
\ifx \showISBNxiii \undefined \def \showISBNxiii  #1{\unskip}     \fi
\ifx \showISSN     \undefined \def \showISSN      #1{\unskip}     \fi
\ifx \showLCCN     \undefined \def \showLCCN      #1{\unskip}     \fi
\ifx \shownote     \undefined \def \shownote      #1{#1}          \fi
\ifx \showarticletitle \undefined \def \showarticletitle #1{#1}   \fi
\ifx \showURL      \undefined \def \showURL       {\relax}        \fi
\providecommand\bibfield[2]{#2}
\providecommand\bibinfo[2]{#2}
\providecommand\natexlab[1]{#1}
\providecommand\showeprint[2][]{arXiv:#2}

\bibitem[Abramova et~al\mbox{.}(2021)]%
        {abramova2021bits}
\bibfield{author}{\bibinfo{person}{Svetlana Abramova}, \bibinfo{person}{Artemij
  Voskobojnikov}, \bibinfo{person}{Konstantin Beznosov}, {and}
  \bibinfo{person}{Rainer B{\"o}hme}.} \bibinfo{year}{2021}\natexlab{}.
\newblock \showarticletitle{Bits under the mattress: Understanding different
  risk perceptions and security behaviors of crypto-asset users}. In
  \bibinfo{booktitle}{\emph{Proceedings of the 2021 CHI Conference on Human
  Factors in Computing Systems}}. \bibinfo{pages}{1--19}.
\newblock


\bibitem[Ackerman and Mainwaring(2005)]%
        {ackerman2005privacy}
\bibfield{author}{\bibinfo{person}{Mark~S Ackerman} {and}
  \bibinfo{person}{Scott~D Mainwaring}.} \bibinfo{year}{2005}\natexlab{}.
\newblock \showarticletitle{Privacy issues and human-computer interaction}.
\newblock \bibinfo{journal}{\emph{Computer}} \bibinfo{volume}{27},
  \bibinfo{number}{5} (\bibinfo{year}{2005}), \bibinfo{pages}{19--26}.
\newblock


\bibitem[Allen(2016)]%
        {allen2016sovereignty}
\bibfield{author}{\bibinfo{person}{Christopher Allen}.}
  \bibinfo{year}{2016}\natexlab{}.
\newblock \bibinfo{title}{The Path to Self-Sovereign Identity}.
\newblock
  \bibinfo{howpublished}{\url{http://www.lifewithalacrity.com/2016/04/the-path-to-self-soverereign-identity.html}}.
\newblock


\bibitem[Ananny and Crawford(2018)]%
        {ananny2018seeing}
\bibfield{author}{\bibinfo{person}{Mike Ananny} {and} \bibinfo{person}{Kate
  Crawford}.} \bibinfo{year}{2018}\natexlab{}.
\newblock \showarticletitle{Seeing without knowing: Limitations of the
  transparency ideal and its application to algorithmic accountability}.
\newblock \bibinfo{journal}{\emph{new media \& society}} \bibinfo{volume}{20},
  \bibinfo{number}{3} (\bibinfo{year}{2018}), \bibinfo{pages}{973--989}.
\newblock


\bibitem[Arafa and Armstrong(2016)]%
        {arafa2016facebook}
\bibfield{author}{\bibinfo{person}{Mohamed Arafa} {and}
  \bibinfo{person}{Crystal Armstrong}.} \bibinfo{year}{2016}\natexlab{}.
\newblock \showarticletitle{"Facebook to mobilize, Twitter to coordinate
  protests, and YouTube to tell the world": New media, cyberactivism, and the
  Arab Spring}.
\newblock \bibinfo{journal}{\emph{Journal of Global Initiatives: Policy,
  Pedagogy, Perspective}} \bibinfo{volume}{10}, \bibinfo{number}{1}
  (\bibinfo{year}{2016}), \bibinfo{pages}{6}.
\newblock


\bibitem[Beikverdi and Song(2015)]%
        {beikverdi2015trend}
\bibfield{author}{\bibinfo{person}{Alireza Beikverdi} {and}
  \bibinfo{person}{JooSeok Song}.} \bibinfo{year}{2015}\natexlab{}.
\newblock \showarticletitle{Trend of centralization in Bitcoin's distributed
  network}. In \bibinfo{booktitle}{\emph{2015 IEEE/ACIS 16th International
  Conference on Software Gngineering, Artificial Intelligence, Networking and
  Parallel/Distributed Computing (SNPD)}}. IEEE, \bibinfo{pages}{1--6}.
\newblock


\bibitem[Berners-Lee et~al\mbox{.}(2001)]%
        {berners2001semantic}
\bibfield{author}{\bibinfo{person}{Tim Berners-Lee}, \bibinfo{person}{James
  Hendler}, {and} \bibinfo{person}{Ora Lassila}.}
  \bibinfo{year}{2001}\natexlab{}.
\newblock \showarticletitle{The semantic web}.
\newblock \bibinfo{journal}{\emph{Scientific American}} \bibinfo{volume}{284},
  \bibinfo{number}{5} (\bibinfo{year}{2001}), \bibinfo{pages}{34--43}.
\newblock


\bibitem[Birhane et~al\mbox{.}(2022)]%
        {birhane2022values}
\bibfield{author}{\bibinfo{person}{Abeba Birhane}, \bibinfo{person}{Pratyusha
  Kalluri}, \bibinfo{person}{Dallas Card}, \bibinfo{person}{William Agnew},
  \bibinfo{person}{Ravit Dotan}, {and} \bibinfo{person}{Michelle Bao}.}
  \bibinfo{year}{2022}\natexlab{}.
\newblock \showarticletitle{The values encoded in machine learning research}.
  In \bibinfo{booktitle}{\emph{2022 ACM Conference on Fairness, Accountability,
  and Transparency}}. \bibinfo{pages}{173--184}.
\newblock


\bibitem[Bod{\'o} et~al\mbox{.}(2021)]%
        {bodo2021decentralisation}
\bibfield{author}{\bibinfo{person}{Bal{\'a}zs Bod{\'o}},
  \bibinfo{person}{Jaya~Klara Brekke}, {and} \bibinfo{person}{Jaap-Henk
  Hoepman}.} \bibinfo{year}{2021}\natexlab{}.
\newblock \showarticletitle{Decentralisation: A multidisciplinary perspective}.
\newblock \bibinfo{journal}{\emph{Internet Policy Review}}
  \bibinfo{volume}{10}, \bibinfo{number}{2} (\bibinfo{year}{2021}),
  \bibinfo{pages}{1--21}.
\newblock


\bibitem[Brekke(2021)]%
        {brekke2021hacker}
\bibfield{author}{\bibinfo{person}{Jaya~Klara Brekke}.}
  \bibinfo{year}{2021}\natexlab{}.
\newblock \showarticletitle{Hacker-engineers and their economies: The political
  economy of decentralised networks and ‘cryptoeconomics’}.
\newblock \bibinfo{journal}{\emph{New Political Economy}} \bibinfo{volume}{26},
  \bibinfo{number}{4} (\bibinfo{year}{2021}), \bibinfo{pages}{646--659}.
\newblock


\bibitem[Brody and Couture(2021)]%
        {brody2021ideologies}
\bibfield{author}{\bibinfo{person}{Ann Brody} {and}
  \bibinfo{person}{St{\'e}phane Couture}.} \bibinfo{year}{2021}\natexlab{}.
\newblock \showarticletitle{Ideologies and imaginaries in blockchain
  communities: The case of ethereum}.
\newblock \bibinfo{journal}{\emph{Canadian Journal of Communication}}
  \bibinfo{volume}{46}, \bibinfo{number}{3} (\bibinfo{year}{2021}),
  \bibinfo{pages}{19--pp}.
\newblock


\bibitem[Buterin(2017)]%
        {buterin2017introduction}
\bibfield{author}{\bibinfo{person}{Vitalik Buterin}.}
  \bibinfo{year}{2017}\natexlab{}.
\newblock \bibinfo{title}{Introduction to cryptoeconomics}.
\newblock \bibinfo{howpublished}{\url{https://www.youtube.com/watch}}.
\newblock


\bibitem[Byrne(2017)]%
        {byrne2017stable}
\bibfield{author}{\bibinfo{person}{Preston Byrne}.}
  \bibinfo{year}{2017}\natexlab{}.
\newblock \bibinfo{title}{Stablecoins are doomed to fail}.
\newblock
  \bibinfo{howpublished}{\url{https://prestonbyrne.com/2017/12/10/stablecoins-are-doomed-to-fail/}}.
\newblock


\bibitem[Caliskan(2022)]%
        {caliskan2022rise}
\bibfield{author}{\bibinfo{person}{Koray Caliskan}.}
  \bibinfo{year}{2022}\natexlab{}.
\newblock \showarticletitle{The rise and fall of Electra: emergence and
  transformation of a global cryptocurrency community}.
\newblock \bibinfo{journal}{\emph{Review of Social Economy}}
  (\bibinfo{year}{2022}), \bibinfo{pages}{1--25}.
\newblock


\bibitem[Chainalysis(2022)]%
        {chainalysiscrime}
\bibfield{author}{\bibinfo{person}{Chainalysis}.}
  \bibinfo{year}{2022}\natexlab{}.
\newblock \bibinfo{title}{The 2023 crypto crime report}.
\newblock
  \bibinfo{howpublished}{\url{https://go.chainalysis.com/rs/503-FAP-074/images/Crypto_Crime_Report_2023.pdf}}.
\newblock
\newblock
\shownote{Accessed: 2023-06-06}.


\bibitem[Charmaz(2014)]%
        {charmaz2014constructing}
\bibfield{author}{\bibinfo{person}{Kathy Charmaz}.}
  \bibinfo{year}{2014}\natexlab{}.
\newblock \bibinfo{booktitle}{\emph{Constructing grounded theory, 2nd ed.}}
\newblock \bibinfo{publisher}{Sage}.
\newblock


\bibitem[Clements(2021)]%
        {clements2021built}
\bibfield{author}{\bibinfo{person}{Ryan Clements}.}
  \bibinfo{year}{2021}\natexlab{}.
\newblock \showarticletitle{Built to fail: The inherent fragility of
  algorithmic stablecoins}.
\newblock \bibinfo{journal}{\emph{Wake Forest Law Review Online}}
  \bibinfo{volume}{11} (\bibinfo{year}{2021}), \bibinfo{pages}{131}.
\newblock


\bibitem[Coinbase(2022)]%
        {coinbaseBTC}
\bibfield{author}{\bibinfo{person}{Coinbase}.} \bibinfo{year}{2022}\natexlab{}.
\newblock \bibinfo{title}{Bitcoin BTC}.
\newblock \bibinfo{howpublished}{\url{https://www.coinbase.com}}.
\newblock
\newblock
\shownote{Accessed: 2022-11-14}.


\bibitem[Cousins et~al\mbox{.}(2019)]%
        {cousins2019value}
\bibfield{author}{\bibinfo{person}{Karlene Cousins}, \bibinfo{person}{Hemang
  Subramanian}, {and} \bibinfo{person}{Pouyan Esmaeilzadeh}.}
  \bibinfo{year}{2019}\natexlab{}.
\newblock \showarticletitle{A value-sensitive design perspective of
  cryptocurrencies: A research agenda}.
\newblock \bibinfo{journal}{\emph{Communications of the association for
  information systems}} \bibinfo{volume}{45}, \bibinfo{number}{1}
  (\bibinfo{year}{2019}), \bibinfo{pages}{27}.
\newblock


\bibitem[Crandall(2019)]%
        {crandall2019blockchains}
\bibfield{author}{\bibinfo{person}{Jillian Crandall}.}
  \bibinfo{year}{2019}\natexlab{}.
\newblock \showarticletitle{Blockchains and the “Chains of Empire”:
  Contextualizing blockchain, cryptocurrency, and neoliberalism in Puerto
  Rico}.
\newblock \bibinfo{journal}{\emph{Design and Culture}} \bibinfo{volume}{11},
  \bibinfo{number}{3} (\bibinfo{year}{2019}), \bibinfo{pages}{279--300}.
\newblock


\bibitem[Cranor and Garfinkel(2005)]%
        {cranor2005security}
\bibfield{author}{\bibinfo{person}{Lorrie~Faith Cranor} {and}
  \bibinfo{person}{Simson Garfinkel}.} \bibinfo{year}{2005}\natexlab{}.
\newblock \bibinfo{booktitle}{\emph{Security and usability: Designing secure
  systems that people can use}}.
\newblock \bibinfo{publisher}{" O'Reilly Media, Inc."}.
\newblock


\bibitem[De(2022)]%
        {DeFTX2022}
\bibfield{author}{\bibinfo{person}{Nikhilesh De}.}
  \bibinfo{year}{2022}\natexlab{}.
\newblock \bibinfo{title}{FTX files for bankruptcy protection in US; CEO
  Bankman-Fried resigns}.
\newblock
\newblock
\urldef\tempurl%
\url{https://www.coindesk.com/policy/2022/11/11/ftx-files-for-bankruptcy-protections-in-us/}
\showURL{%
\tempurl}


\bibitem[De~Filippi and Loveluck(2016)]%
        {defillipi2016invisible}
\bibfield{author}{\bibinfo{person}{Primavera De~Filippi} {and}
  \bibinfo{person}{Benjamin Loveluck}.} \bibinfo{year}{2016}\natexlab{}.
\newblock \showarticletitle{The invisible politics of bitcoin: Governance
  crisis of a decentralized infrastructure}.
\newblock \bibinfo{journal}{\emph{Internet Policy Review}} \bibinfo{volume}{5},
  \bibinfo{number}{4} (\bibinfo{year}{2016}).
\newblock


\bibitem[Dodd(2018)]%
        {dodd2018social}
\bibfield{author}{\bibinfo{person}{Nigel Dodd}.}
  \bibinfo{year}{2018}\natexlab{}.
\newblock \showarticletitle{The social life of Bitcoin}.
\newblock \bibinfo{journal}{\emph{Theory, Culture \& Society}}
  \bibinfo{volume}{35}, \bibinfo{number}{3} (\bibinfo{year}{2018}),
  \bibinfo{pages}{35--56}.
\newblock


\bibitem[Dotan and Milli(2019)]%
        {dotan2019value}
\bibfield{author}{\bibinfo{person}{Ravit Dotan} {and} \bibinfo{person}{Smitha
  Milli}.} \bibinfo{year}{2019}\natexlab{}.
\newblock \showarticletitle{Value-laden disciplinary shifts in machine
  learning}.
\newblock \bibinfo{journal}{\emph{arXiv preprint arXiv:1912.01172}}
  (\bibinfo{year}{2019}).
\newblock


\bibitem[DuPont(2017)]%
        {dupont2017experiments}
\bibfield{author}{\bibinfo{person}{Quinn DuPont}.}
  \bibinfo{year}{2017}\natexlab{}.
\newblock \showarticletitle{Experiments in algorithmic governance: A history
  and ethnography of “The DAO,” a failed decentralized autonomous
  organization}.
\newblock In \bibinfo{booktitle}{\emph{Bitcoin and Beyond}}.
  \bibinfo{publisher}{Routledge}, \bibinfo{pages}{157--177}.
\newblock


\bibitem[Elsden et~al\mbox{.}(2018)]%
        {elsden2018making}
\bibfield{author}{\bibinfo{person}{Chris Elsden}, \bibinfo{person}{Arthi
  Manohar}, \bibinfo{person}{Jo Briggs}, \bibinfo{person}{Mike Harding},
  \bibinfo{person}{Chris Speed}, {and} \bibinfo{person}{John Vines}.}
  \bibinfo{year}{2018}\natexlab{}.
\newblock \showarticletitle{Making sense of blockchain applications: A typology
  for HCI}. In \bibinfo{booktitle}{\emph{Proceedings of the 2018 CHI Conference
  on Human Factors in Computing Systems}}. \bibinfo{pages}{1--14}.
\newblock


\bibitem[Fan et~al\mbox{.}(2023)]%
        {fan2023altruistic}
\bibfield{author}{\bibinfo{person}{Sizheng Fan}, \bibinfo{person}{Tian Min},
  \bibinfo{person}{Xiao Wu}, {and} \bibinfo{person}{Wei Cai}.}
  \bibinfo{year}{2023}\natexlab{}.
\newblock \showarticletitle{Altruistic and profit-oriented: Making sense of
  roles in Web3 community from airdrop perspective}. In
  \bibinfo{booktitle}{\emph{Proceedings of the 2023 CHI Conference on Human
  Factors in Computing Systems}}. \bibinfo{pages}{1--16}.
\newblock


\bibitem[Fard~Bahreini et~al\mbox{.}(2021)]%
        {fard2021distributing}
\bibfield{author}{\bibinfo{person}{Amir Fard~Bahreini}, \bibinfo{person}{John
  Collomosse}, \bibinfo{person}{Marc-David~L Seidel}, \bibinfo{person}{Maral
  Sotoudehnia}, {and} \bibinfo{person}{Carson~C Woo}.}
  \bibinfo{year}{2021}\natexlab{}.
\newblock \showarticletitle{Distributing and democratizing institutional power
  through decentralization}.
\newblock In \bibinfo{booktitle}{\emph{Building Decentralized Trust}}.
  \bibinfo{publisher}{Springer}, \bibinfo{pages}{95--109}.
\newblock


\bibitem[Fehr and Camerer(2007)]%
        {fehr2007social}
\bibfield{author}{\bibinfo{person}{Ernst Fehr} {and} \bibinfo{person}{Colin~F
  Camerer}.} \bibinfo{year}{2007}\natexlab{}.
\newblock \showarticletitle{Social neuroeconomics: The neural circuitry of
  social preferences}.
\newblock \bibinfo{journal}{\emph{Trends in Cognitive Sciences}}
  \bibinfo{volume}{11}, \bibinfo{number}{10} (\bibinfo{year}{2007}),
  \bibinfo{pages}{419--427}.
\newblock


\bibitem[Friedman and Hendry(2019)]%
        {friedman2019value}
\bibfield{author}{\bibinfo{person}{Batya Friedman} {and}
  \bibinfo{person}{David~G Hendry}.} \bibinfo{year}{2019}\natexlab{}.
\newblock \bibinfo{booktitle}{\emph{Value sensitive design: Shaping technology
  with moral imagination}}.
\newblock \bibinfo{publisher}{MIT Press}.
\newblock


\bibitem[Friedman et~al\mbox{.}(2013)]%
        {friedman2013value}
\bibfield{author}{\bibinfo{person}{Batya Friedman}, \bibinfo{person}{Peter~H
  Kahn}, \bibinfo{person}{Alan Borning}, {and} \bibinfo{person}{Alina
  Huldtgren}.} \bibinfo{year}{2013}\natexlab{}.
\newblock \showarticletitle{Value sensitive design and information systems}.
\newblock \bibinfo{journal}{\emph{Early engagement and new technologies:
  Opening up the laboratory}} (\bibinfo{year}{2013}), \bibinfo{pages}{55--95}.
\newblock


\bibitem[Fr{\"o}hlich et~al\mbox{.}(2020)]%
        {frohlich2020don}
\bibfield{author}{\bibinfo{person}{Michael Fr{\"o}hlich},
  \bibinfo{person}{Felix Gutjahr}, {and} \bibinfo{person}{Florian Alt}.}
  \bibinfo{year}{2020}\natexlab{}.
\newblock \showarticletitle{Don't lose your coin! Investigating Security
  practices of cryptocurrency users}. In \bibinfo{booktitle}{\emph{Proceedings
  of the 2020 ACM Designing Interactive Systems Conference}}.
  \bibinfo{pages}{1751--1763}.
\newblock


\bibitem[Gao et~al\mbox{.}(2016)]%
        {gao2016two}
\bibfield{author}{\bibinfo{person}{Xianyi Gao}, \bibinfo{person}{Gradeigh~D
  Clark}, {and} \bibinfo{person}{Janne Lindqvist}.}
  \bibinfo{year}{2016}\natexlab{}.
\newblock \showarticletitle{Of two minds, multiple addresses, and one ledger:
  characterizing opinions, knowledge, and perceptions of Bitcoin across users
  and non-users}. In \bibinfo{booktitle}{\emph{Proceedings of the 2016 CHI
  conference on human factors in computing systems}}.
  \bibinfo{pages}{1656--1668}.
\newblock


\bibitem[Gencer et~al\mbox{.}(2018)]%
        {gencer2018decentralization}
\bibfield{author}{\bibinfo{person}{Adem~Efe Gencer}, \bibinfo{person}{Soumya
  Basu}, \bibinfo{person}{Ittay Eyal}, \bibinfo{person}{Robbert Van~Renesse},
  {and} \bibinfo{person}{Emin~G{\"u}n Sirer}.} \bibinfo{year}{2018}\natexlab{}.
\newblock \showarticletitle{Decentralization in Bitcoin and Ethereum networks}.
  In \bibinfo{booktitle}{\emph{Financial Cryptography and Data Security: 22nd
  International Conference, FC 2018, Nieuwpoort, Cura{\c{c}}ao, February
  26--March 2, 2018, Revised Selected Papers 22}}. Springer,
  \bibinfo{pages}{439--457}.
\newblock


\bibitem[Gladstone et~al\mbox{.}(2022)]%
        {gladstone2022celsius}
\bibfield{author}{\bibinfo{person}{Alexander Gladstone},
  \bibinfo{person}{Vicky~Ge Huang}, {and} \bibinfo{person}{Soma Biswas}.}
  \bibinfo{year}{2022}\natexlab{}.
\newblock \showarticletitle{Crypto crash drags lender Celsius Network into
  bankruptcy}.
\newblock \bibinfo{journal}{\emph{The Wall Street Journal}}
  (\bibinfo{date}{July} \bibinfo{year}{2022}).
\newblock
\urldef\tempurl%
\url{https://www.wsj.com/articles/crypto-crash-drags-lender-celsius-network-into-bankruptcy-11657758483}
\showURL{%
\tempurl}


\bibitem[Glaser(1978)]%
        {glaser1978theoretical}
\bibfield{author}{\bibinfo{person}{Barney~G Glaser}.}
  \bibinfo{year}{1978}\natexlab{}.
\newblock \bibinfo{booktitle}{\emph{Theoretical sensitivity}}.
\newblock \bibinfo{publisher}{University of California,}.
\newblock


\bibitem[Glaser et~al\mbox{.}(1968)]%
        {glaser1968discovery}
\bibfield{author}{\bibinfo{person}{Barney~G Glaser}, \bibinfo{person}{Anselm~L
  Strauss}, {and} \bibinfo{person}{Elizabeth Strutzel}.}
  \bibinfo{year}{1968}\natexlab{}.
\newblock \showarticletitle{The discovery of grounded theory: Strategies for
  qualitative research}.
\newblock \bibinfo{journal}{\emph{Nursing research}} \bibinfo{volume}{17},
  \bibinfo{number}{4} (\bibinfo{year}{1968}), \bibinfo{pages}{364}.
\newblock


\bibitem[Golumbia(2016)]%
        {golumbia2016politics}
\bibfield{author}{\bibinfo{person}{David Golumbia}.}
  \bibinfo{year}{2016}\natexlab{}.
\newblock \bibinfo{booktitle}{\emph{The politics of Bitcoin: Software as
  right-wing extremism}}.
\newblock \bibinfo{publisher}{University of Minnesota Press}.
\newblock


\bibitem[Hassan et~al\mbox{.}(2018)]%
        {hassan2018exposure}
\bibfield{author}{\bibinfo{person}{Ghayda Hassan},
  \bibinfo{person}{S{\'e}bastien Brouillette-Alarie},
  \bibinfo{person}{S{\'e}raphin Alava}, \bibinfo{person}{Divina Frau-Meigs},
  \bibinfo{person}{Lysiane Lavoie}, \bibinfo{person}{Arber Fetiu},
  \bibinfo{person}{Wynnpaul Varela}, \bibinfo{person}{Evgueni Borokhovski},
  \bibinfo{person}{Vivek Venkatesh}, \bibinfo{person}{C{\'e}cile Rousseau},
  {et~al\mbox{.}}} \bibinfo{year}{2018}\natexlab{}.
\newblock \showarticletitle{Exposure to extremist online content could lead to
  violent radicalization: A systematic review of empirical evidence}.
\newblock \bibinfo{journal}{\emph{International Journal of Developmental
  Science}} \bibinfo{volume}{12}, \bibinfo{number}{1-2} (\bibinfo{year}{2018}),
  \bibinfo{pages}{71--88}.
\newblock


\bibitem[Hayes(2019)]%
        {hayes2019socio}
\bibfield{author}{\bibinfo{person}{Adam Hayes}.}
  \bibinfo{year}{2019}\natexlab{}.
\newblock \showarticletitle{The socio-technological lives of bitcoin}.
\newblock \bibinfo{journal}{\emph{Theory, Culture \& Society}}
  \bibinfo{volume}{36}, \bibinfo{number}{4} (\bibinfo{year}{2019}),
  \bibinfo{pages}{49--72}.
\newblock


\bibitem[Herian(2018)]%
        {herian2018blockchain}
\bibfield{author}{\bibinfo{person}{Robert Herian}.}
  \bibinfo{year}{2018}\natexlab{}.
\newblock \showarticletitle{Blockchain and the distributed reproduction of
  capitalist class power}.
\newblock \bibinfo{journal}{\emph{MoneyLab Reader}}  \bibinfo{volume}{2}
  (\bibinfo{year}{2018}), \bibinfo{pages}{43--51}.
\newblock


\bibitem[Higgins(2018)]%
        {higgins_sec_2018}
\bibfield{author}{\bibinfo{person}{Stan Higgins}.}
  \bibinfo{year}{2018}\natexlab{}.
\newblock \bibinfo{title}{{SEC} {Chief} {Clayton}: '{Every} {ICO} {I}'ve {Seen}
  {is} a {security}'}.
\newblock
\newblock
\urldef\tempurl%
\url{https://www.coindesk.com/markets/2018/02/06/sec-chief-clayton-every-ico-ive-seen-is-a-security/}
\showURL{%
\tempurl}


\bibitem[Hinman(2018)]%
        {hinman2018}
\bibfield{author}{\bibinfo{person}{William Hinman}.}
  \bibinfo{year}{2018}\natexlab{}.
\newblock \bibinfo{title}{Digital asset transactions: When Howey met Gary
  (plastic)}.
\newblock
\newblock
\urldef\tempurl%
\url{https://www.sec.gov/news/speech/speech-hinman-061418,}
\showURL{%
\tempurl}
\newblock
\shownote{Remarks at the Yahoo Finance All Markets Summit: Crypto[Accessed:
  2023 05 24]}.


\bibitem[Hope et~al\mbox{.}(2019)]%
        {hope2019hackathons}
\bibfield{author}{\bibinfo{person}{Alexis Hope}, \bibinfo{person}{Catherine
  D'Ignazio}, \bibinfo{person}{Josephine Hoy}, \bibinfo{person}{Rebecca
  Michelson}, \bibinfo{person}{Jennifer Roberts}, \bibinfo{person}{Kate
  Krontiris}, {and} \bibinfo{person}{Ethan Zuckerman}.}
  \bibinfo{year}{2019}\natexlab{}.
\newblock \showarticletitle{Hackathons as participatory design: iterating
  feminist utopias}. In \bibinfo{booktitle}{\emph{Proceedings of the 2019 CHI
  Conference on Human Factors in Computing Systems}}. \bibinfo{pages}{1--14}.
\newblock


\bibitem[Hoyng(2023)]%
        {hoyng2023bitcoin}
\bibfield{author}{\bibinfo{person}{Rolien Hoyng}.}
  \bibinfo{year}{2023}\natexlab{}.
\newblock \showarticletitle{From Bitcoin to farm bank: An idiotic inquiry into
  blockchain speculation}.
\newblock \bibinfo{journal}{\emph{Convergence}} (\bibinfo{year}{2023}),
  \bibinfo{pages}{13548565231154104}.
\newblock


\bibitem[Husain et~al\mbox{.}(2020)]%
        {husain2020political}
\bibfield{author}{\bibinfo{person}{Syed~Omer Husain}, \bibinfo{person}{Alex
  Franklin}, {and} \bibinfo{person}{Dirk Roep}.}
  \bibinfo{year}{2020}\natexlab{}.
\newblock \showarticletitle{The political imaginaries of blockchain projects:
  Discerning the expressions of an emerging ecosystem}.
\newblock \bibinfo{journal}{\emph{Sustainability Science}}
  \bibinfo{volume}{15}, \bibinfo{number}{2} (\bibinfo{year}{2020}),
  \bibinfo{pages}{379--394}.
\newblock


\bibitem[H{\"u}tten(2019)]%
        {hutten2019soft}
\bibfield{author}{\bibinfo{person}{Moritz H{\"u}tten}.}
  \bibinfo{year}{2019}\natexlab{}.
\newblock \showarticletitle{The soft spot of hard code: blockchain technology,
  network governance and pitfalls of technological utopianism}.
\newblock \bibinfo{journal}{\emph{Global Networks}} \bibinfo{volume}{19},
  \bibinfo{number}{3} (\bibinfo{year}{2019}), \bibinfo{pages}{329--348}.
\newblock


\bibitem[Iachello et~al\mbox{.}(2007)]%
        {iachello2007end}
\bibfield{author}{\bibinfo{person}{Giovanni Iachello}, \bibinfo{person}{Jason
  Hong}, {et~al\mbox{.}}} \bibinfo{year}{2007}\natexlab{}.
\newblock \showarticletitle{End-user privacy in human--computer interaction}.
\newblock \bibinfo{journal}{\emph{Foundations and Trends{\textregistered} in
  Human--Computer Interaction}} \bibinfo{volume}{1}, \bibinfo{number}{1}
  (\bibinfo{year}{2007}), \bibinfo{pages}{1--137}.
\newblock


\bibitem[Inwood and Zappavigna(2021)]%
        {inwood2021ideology}
\bibfield{author}{\bibinfo{person}{Olivia Inwood} {and}
  \bibinfo{person}{Michele Zappavigna}.} \bibinfo{year}{2021}\natexlab{}.
\newblock \showarticletitle{Ideology, attitudinal positioning, and the
  blockchain: A social semiotic approach to understanding the values construed
  in the whitepapers of blockchain start-ups}.
\newblock \bibinfo{journal}{\emph{Social Semiotics}} (\bibinfo{year}{2021}),
  \bibinfo{pages}{1--19}.
\newblock


\bibitem[Jabbar and Bj{\o}rn(2019)]%
        {jabbar2019blockchain}
\bibfield{author}{\bibinfo{person}{Karim Jabbar} {and}
  \bibinfo{person}{Pernille Bj{\o}rn}.} \bibinfo{year}{2019}\natexlab{}.
\newblock \showarticletitle{Blockchain assemblages: Whiteboxing technology and
  transforming infrastructural imaginaries}. In
  \bibinfo{booktitle}{\emph{Proceedings of the 2019 CHI Conference on Human
  Factors in Computing Systems}}. \bibinfo{pages}{1--13}.
\newblock


\bibitem[Jerolmack and Khan(2014)]%
        {jerolmack2014talk}
\bibfield{author}{\bibinfo{person}{Colin Jerolmack} {and}
  \bibinfo{person}{Shamus Khan}.} \bibinfo{year}{2014}\natexlab{}.
\newblock \showarticletitle{Talk is cheap: Ethnography and the attitudinal
  fallacy}.
\newblock \bibinfo{journal}{\emph{Sociological Methods \& Research}}
  \bibinfo{volume}{43}, \bibinfo{number}{2} (\bibinfo{year}{2014}),
  \bibinfo{pages}{178--209}.
\newblock


\bibitem[Kahneman and Tversky(1979)]%
        {kahneman1979prospect}
\bibfield{author}{\bibinfo{person}{Daniel Kahneman} {and} \bibinfo{person}{Amos
  Tversky}.} \bibinfo{year}{1979}\natexlab{}.
\newblock \showarticletitle{Prospect theory: An analysis of decision under
  risk}.
\newblock \bibinfo{journal}{\emph{Econometrica}} \bibinfo{volume}{47},
  \bibinfo{number}{2} (\bibinfo{year}{1979}), \bibinfo{pages}{263--292}.
\newblock


\bibitem[Kapiszewski and Karcher(2021)]%
        {kapiszewski2021transparency}
\bibfield{author}{\bibinfo{person}{Diana Kapiszewski} {and}
  \bibinfo{person}{Sebastian Karcher}.} \bibinfo{year}{2021}\natexlab{}.
\newblock \showarticletitle{Transparency in practice in qualitative research}.
\newblock \bibinfo{journal}{\emph{Political Science \& Politics}}
  \bibinfo{volume}{54}, \bibinfo{number}{2} (\bibinfo{year}{2021}),
  \bibinfo{pages}{285--291}.
\newblock


\bibitem[Knittel et~al\mbox{.}(2019)]%
        {knittel2019most}
\bibfield{author}{\bibinfo{person}{Megan Knittel}, \bibinfo{person}{Shelby
  Pitts}, {and} \bibinfo{person}{Rick Wash}.} \bibinfo{year}{2019}\natexlab{}.
\newblock \showarticletitle{"The most trustworthy coin": How ideological
  tensions drive trust in Bitcoin}.
\newblock \bibinfo{journal}{\emph{Proceedings of the ACM on Human-Computer
  Interaction}} \bibinfo{volume}{3}, \bibinfo{number}{CSCW}
  (\bibinfo{year}{2019}), \bibinfo{pages}{1--23}.
\newblock


\bibitem[Krisztian~Sandor(2022)]%
        {sandorTerra2022}
\bibfield{author}{\bibinfo{person}{Ekin~Genç Krisztian~Sandor}.}
  \bibinfo{year}{2022}\natexlab{}.
\newblock \bibinfo{title}{The fall of Terra: A timeline of the meteoric rise
  and crash of UST and LUNA}.
\newblock
\newblock
\urldef\tempurl%
\url{https://www.coindesk.com/learn/the-fall-of-terra-a-timeline-of-the-meteoric-rise-and-crash-of-ust-and-luna/}
\showURL{%
\tempurl}


\bibitem[Krombholz et~al\mbox{.}(2017)]%
        {krombholz2017other}
\bibfield{author}{\bibinfo{person}{Katharina Krombholz},
  \bibinfo{person}{Aljosha Judmayer}, \bibinfo{person}{Matthias Gusenbauer},
  {and} \bibinfo{person}{Edgar Weippl}.} \bibinfo{year}{2017}\natexlab{}.
\newblock \showarticletitle{The other side of the coin: User experiences with
  bitcoin security and privacy}. In \bibinfo{booktitle}{\emph{Financial
  Cryptography and Data Security: 20th International Conference, FC 2016,
  Christ Church, Barbados, February 22--26, 2016, Revised Selected Papers 20}}.
  Springer, \bibinfo{pages}{555--580}.
\newblock


\bibitem[Kuhn(2021)]%
        {kuhn2021HFSP}
\bibfield{author}{\bibinfo{person}{Daniel Kuhn}.}
  \bibinfo{year}{2021}\natexlab{}.
\newblock \bibinfo{title}{The Decoder: "Have fun staying poor"}.
\newblock
  \bibinfo{howpublished}{\url{https://www.coindesk.com/markets/2021/03/03/the-decoder-have-fun-staying-poor}}.
\newblock


\bibitem[Laatikainen et~al\mbox{.}(2021)]%
        {laatikainen2021towards}
\bibfield{author}{\bibinfo{person}{Gabriella Laatikainen},
  \bibinfo{person}{Taija Kolehmainen}, \bibinfo{person}{Mengcheng Li},
  \bibinfo{person}{Markus Hautala}, \bibinfo{person}{Antti Kettunen}, {and}
  \bibinfo{person}{Pekka Abrahamsson}.} \bibinfo{year}{2021}\natexlab{}.
\newblock \showarticletitle{Towards a trustful digital World: Exploring
  self-Sovereign identity ecosystems}. In \bibinfo{booktitle}{\emph{Pacific
  Asia Conference on Information Systems}}. Association for Information
  Systems.
\newblock


\bibitem[Lee et~al\mbox{.}(2023)]%
        {lee2023social}
\bibfield{author}{\bibinfo{person}{Jiyoung Lee}, \bibinfo{person}{Jihyang
  Choi}, {and} \bibinfo{person}{Rebecca~K Britt}.}
  \bibinfo{year}{2023}\natexlab{}.
\newblock \showarticletitle{Social media as risk-attenuation and
  misinformation-amplification station: How social media interaction affects
  misperceptions about COVID-19}.
\newblock \bibinfo{journal}{\emph{Health communication}} \bibinfo{volume}{38},
  \bibinfo{number}{6} (\bibinfo{year}{2023}), \bibinfo{pages}{1232--1242}.
\newblock


\bibitem[Lee et~al\mbox{.}(2022)]%
        {lee2022password}
\bibfield{author}{\bibinfo{person}{Kevin Lee}, \bibinfo{person}{Sten
  Sj{\"o}berg}, {and} \bibinfo{person}{Arvind Narayanan}.}
  \bibinfo{year}{2022}\natexlab{}.
\newblock \showarticletitle{Password policies of most top websites fail to
  follow best practices}. In \bibinfo{booktitle}{\emph{Eighteenth Symposium on
  Usable Privacy and Security (SOUPS 2022)}}. \bibinfo{pages}{561--580}.
\newblock


\bibitem[Lenhard(2021)]%
        {lenhard2021}
\bibfield{author}{\bibinfo{person}{Johannes Lenhard}.}
  \bibinfo{year}{2021}\natexlab{}.
\newblock \bibinfo{booktitle}{\emph{Work, Society, and the Ethical Self:
  Chimeras of Freedom in the Neoliberal Era}}.
\newblock \bibinfo{publisher}{Berghahn Books}, Chapter Unicorn-makers working
  for freedom (and monopolies), \bibinfo{pages}{221--238}.
\newblock


\bibitem[Liaqat et~al\mbox{.}(2021)]%
        {liaqat2021participatory}
\bibfield{author}{\bibinfo{person}{Amna Liaqat}, \bibinfo{person}{Benett
  Axtell}, {and} \bibinfo{person}{Cosmin Munteanu}.}
  \bibinfo{year}{2021}\natexlab{}.
\newblock \showarticletitle{Participatory design for intergenerational culture
  exchange in immigrant families: How collaborative narration and creation
  fosters democratic engagement}.
\newblock \bibinfo{journal}{\emph{Proceedings of the ACM on Human-Computer
  Interaction}} \bibinfo{volume}{5}, \bibinfo{number}{CSCW1}
  (\bibinfo{year}{2021}), \bibinfo{pages}{1--40}.
\newblock


\bibitem[Lustig(2019)]%
        {lustig2019intersecting}
\bibfield{author}{\bibinfo{person}{Caitlin Lustig}.}
  \bibinfo{year}{2019}\natexlab{}.
\newblock \showarticletitle{Intersecting imaginaries: Visions of decentralized
  autonomous systems}.
\newblock \bibinfo{journal}{\emph{Proceedings of the ACM on Human-Computer
  Interaction}} \bibinfo{volume}{3}, \bibinfo{number}{CSCW}
  (\bibinfo{year}{2019}), \bibinfo{pages}{1--27}.
\newblock


\bibitem[Mai et~al\mbox{.}(2020)]%
        {mai2020user}
\bibfield{author}{\bibinfo{person}{Alexandra Mai}, \bibinfo{person}{Katharina
  Pfeffer}, \bibinfo{person}{Matthias Gusenbauer}, \bibinfo{person}{Edgar
  Weippl}, {and} \bibinfo{person}{Katharina Krombholz}.}
  \bibinfo{year}{2020}\natexlab{}.
\newblock \showarticletitle{User mental models of cryptocurrency systems—A
  grounded theory approach}.
\newblock  (\bibinfo{year}{2020}).
\newblock


\bibitem[Mattke et~al\mbox{.}(2021)]%
        {mattke2021bitcoin}
\bibfield{author}{\bibinfo{person}{Jens Mattke}, \bibinfo{person}{Christian
  Maier}, \bibinfo{person}{Lea Reis}, {and} \bibinfo{person}{Tim Weitzel}.}
  \bibinfo{year}{2021}\natexlab{}.
\newblock \showarticletitle{Bitcoin investment: A mixed-methods study of
  investment motivations}.
\newblock \bibinfo{journal}{\emph{European Journal of Information Systems}}
  \bibinfo{volume}{30}, \bibinfo{number}{3} (\bibinfo{year}{2021}),
  \bibinfo{pages}{261--285}.
\newblock


\bibitem[Maurer et~al\mbox{.}(2013)]%
        {maurer2013perhaps}
\bibfield{author}{\bibinfo{person}{Bill Maurer}, \bibinfo{person}{Taylor~C
  Nelms}, {and} \bibinfo{person}{Lana Swartz}.}
  \bibinfo{year}{2013}\natexlab{}.
\newblock \showarticletitle{“When perhaps the real problem is money
  itself!”: The practical materiality of Bitcoin}.
\newblock \bibinfo{journal}{\emph{Social Semiotics}} \bibinfo{volume}{23},
  \bibinfo{number}{2} (\bibinfo{year}{2013}), \bibinfo{pages}{261--277}.
\newblock


\bibitem[McDonald et~al\mbox{.}(2019)]%
        {mcdonald2019reliability}
\bibfield{author}{\bibinfo{person}{Nora McDonald}, \bibinfo{person}{Sarita
  Schoenebeck}, {and} \bibinfo{person}{Andrea Forte}.}
  \bibinfo{year}{2019}\natexlab{}.
\newblock \showarticletitle{Reliability and inter-rater reliability in
  qualitative research: Norms and guidelines for CSCW and HCI practice}.
\newblock \bibinfo{journal}{\emph{Proceedings of the ACM on Human-Computer
  Interaction}} \bibinfo{volume}{3}, \bibinfo{number}{CSCW}
  (\bibinfo{year}{2019}), \bibinfo{pages}{1--23}.
\newblock


\bibitem[Miller(2022)]%
        {miller2022terra}
\bibfield{author}{\bibinfo{person}{Hannah Miller}.}
  \bibinfo{year}{2022}\natexlab{}.
\newblock \bibinfo{title}{Crypto investing and the curse of the Luna tattoo}.
\newblock
  \bibinfo{howpublished}{\url{https://www.bloomberg.com/news/newsletters/2022-06-02/crypto-investing-mike-novogratz-and-the-curse-of-the-luna-tattoo}}.
\newblock


\bibitem[Mirowski(2014)]%
        {mirowski2014never}
\bibfield{author}{\bibinfo{person}{Philip Mirowski}.}
  \bibinfo{year}{2014}\natexlab{}.
\newblock \bibinfo{booktitle}{\emph{Never let a serious crisis go to waste: How
  neoliberalism survived the financial meltdown}}.
\newblock \bibinfo{publisher}{Verso Books}.
\newblock


\bibitem[Morris(2023a)]%
        {morriscryptocrooks}
\bibfield{author}{\bibinfo{person}{David~Z. Morris}.}
  \bibinfo{year}{2023}\natexlab{a}.
\newblock \bibinfo{title}{Lunacy episode 2: Confidence game}.
\newblock
  \bibinfo{howpublished}{\url{https://www.coindesk.com/podcasts/crypto-crooks/lunacy-episode-2-confidence-game/}}.
\newblock


\bibitem[Morris(2023b)]%
        {morriscryptocrooks5}
\bibfield{author}{\bibinfo{person}{David~Z. Morris}.}
  \bibinfo{year}{2023}\natexlab{b}.
\newblock \bibinfo{title}{Lunacy episode 5: Between the moon and
  Montenegro–Do Kwon goes to jail}.
\newblock
  \bibinfo{howpublished}{\url{https://www.coindesk.com/podcasts/crypto-crooks/lunacy-episode-5-between-the-moon-and-montenegro-do-kwon-goes-to-jail/}}.
\newblock


\bibitem[Mundt et~al\mbox{.}(2018)]%
        {mundt2018scaling}
\bibfield{author}{\bibinfo{person}{Marcia Mundt}, \bibinfo{person}{Karen Ross},
  {and} \bibinfo{person}{Charla~M Burnett}.} \bibinfo{year}{2018}\natexlab{}.
\newblock \showarticletitle{Scaling social movements through social media: The
  case of Black Lives Matter}.
\newblock \bibinfo{journal}{\emph{Social Media+ Society}} \bibinfo{volume}{4},
  \bibinfo{number}{4} (\bibinfo{year}{2018}),
  \bibinfo{pages}{2056305118807911}.
\newblock


\bibitem[Murphy et~al\mbox{.}(2021)]%
        {murphy2021ethnography}
\bibfield{author}{\bibinfo{person}{Alexandra~K Murphy}, \bibinfo{person}{Colin
  Jerolmack}, {and} \bibinfo{person}{DeAnna Smith}.}
  \bibinfo{year}{2021}\natexlab{}.
\newblock \showarticletitle{Ethnography, data transparency, and the information
  age}.
\newblock \bibinfo{journal}{\emph{Annual Review of Sociology}}
  \bibinfo{volume}{47} (\bibinfo{year}{2021}), \bibinfo{pages}{41--61}.
\newblock


\bibitem[Nakamoto(2008)]%
        {nakamoto2008bitcoin}
\bibfield{author}{\bibinfo{person}{Satoshi Nakamoto}.}
  \bibinfo{year}{2008}\natexlab{}.
\newblock \showarticletitle{Bitcoin: A peer-to-peer electronic cash system}.
\newblock \bibinfo{journal}{\emph{Decentralized Business Review}}
  (\bibinfo{year}{2008}), \bibinfo{pages}{21260}.
\newblock


\bibitem[Napolitano(2022)]%
        {NapolitanoCelsius2022}
\bibfield{author}{\bibinfo{person}{Elizabeth Napolitano}.}
  \bibinfo{year}{2022}\natexlab{}.
\newblock \bibinfo{title}{The fall of Celsius Network: A timeline of the crypto
  lender’s descent into insolvency}.
\newblock
\newblock
\urldef\tempurl%
\url{https://www.coindesk.com/markets/2022/07/15/the-fall-of-celsius-network-a-timeline-of-the-crypto-lenders-descent-into-insolvency/}
\showURL{%
\tempurl}


\bibitem[Narayanan(2013)]%
        {narayanan2013happened}
\bibfield{author}{\bibinfo{person}{Arvind Narayanan}.}
  \bibinfo{year}{2013}\natexlab{}.
\newblock \showarticletitle{What happened to the crypto dream?, part 1}.
\newblock \bibinfo{journal}{\emph{IEEE Security \& Privacy}}
  \bibinfo{volume}{11}, \bibinfo{number}{2} (\bibinfo{year}{2013}),
  \bibinfo{pages}{75--76}.
\newblock


\bibitem[Narayanan et~al\mbox{.}(2016)]%
        {narayanan2016bitcoin}
\bibfield{author}{\bibinfo{person}{Arvind Narayanan}, \bibinfo{person}{Joseph
  Bonneau}, \bibinfo{person}{Edward Felten}, \bibinfo{person}{Andrew Miller},
  {and} \bibinfo{person}{Steven Goldfeder}.} \bibinfo{year}{2016}\natexlab{}.
\newblock \bibinfo{booktitle}{\emph{Bitcoin and cryptocurrency technologies: A
  comprehensive introduction}}.
\newblock \bibinfo{publisher}{Princeton University Press}.
\newblock


\bibitem[Narayanan and Clark(2017)]%
        {narayanan2017bitcoin}
\bibfield{author}{\bibinfo{person}{Arvind Narayanan} {and}
  \bibinfo{person}{Jeremy Clark}.} \bibinfo{year}{2017}\natexlab{}.
\newblock \showarticletitle{Bitcoin's academic pedigree}.
\newblock \bibinfo{journal}{\emph{Commun. ACM}} \bibinfo{volume}{60},
  \bibinfo{number}{12} (\bibinfo{year}{2017}), \bibinfo{pages}{36--45}.
\newblock


\bibitem[Nicholls et~al\mbox{.}(2006)]%
        {nicholls2006satisfaction}
\bibfield{author}{\bibinfo{person}{Michael~ER Nicholls},
  \bibinfo{person}{Catherine~A Orr}, \bibinfo{person}{Matia Okubo}, {and}
  \bibinfo{person}{Andrea Loftus}.} \bibinfo{year}{2006}\natexlab{}.
\newblock \showarticletitle{Satisfaction guaranteed: The effect of spatial
  biases on responses to Likert scales}.
\newblock \bibinfo{journal}{\emph{Psychological Science}} \bibinfo{volume}{17},
  \bibinfo{number}{12} (\bibinfo{year}{2006}), \bibinfo{pages}{1027--1028}.
\newblock


\bibitem[Oliver(1991)]%
        {oliver1991strategic}
\bibfield{author}{\bibinfo{person}{Christine Oliver}.}
  \bibinfo{year}{1991}\natexlab{}.
\newblock \showarticletitle{Strategic responses to institutional processes}.
\newblock \bibinfo{journal}{\emph{Academy of Management Review}}
  \bibinfo{volume}{16}, \bibinfo{number}{1} (\bibinfo{year}{1991}),
  \bibinfo{pages}{145--179}.
\newblock


\bibitem[Ostroff(2022)]%
        {ostroff2022terra}
\bibfield{author}{\bibinfo{person}{Caitlin Ostroff}.}
  \bibinfo{year}{2022}\natexlab{}.
\newblock \showarticletitle{Why did cryptocurrencies TerraUSD and Luna unravel?
  Stablecoin price crash explained}.
\newblock \bibinfo{journal}{\emph{The Wall Street Journal}}
  (\bibinfo{date}{May} \bibinfo{year}{2022}).
\newblock
\urldef\tempurl%
\url{https://www.wsj.com/articles/why-did-cryptocurrencies-terrausd-and-luna-unravel-stablecoin-price-crash-explained-11652462779}
\showURL{%
\tempurl}


\bibitem[Pillow(2003)]%
        {pillow2003confession}
\bibfield{author}{\bibinfo{person}{Wanda Pillow}.}
  \bibinfo{year}{2003}\natexlab{}.
\newblock \showarticletitle{Confession, catharsis, or cure? Rethinking the uses
  of reflexivity as methodological power in qualitative research}.
\newblock \bibinfo{journal}{\emph{International Journal of Qualitative Studies
  in Education}} \bibinfo{volume}{16}, \bibinfo{number}{2}
  (\bibinfo{year}{2003}), \bibinfo{pages}{175--196}.
\newblock


\bibitem[Podsakoff et~al\mbox{.}(2003)]%
        {podsakoff2003common}
\bibfield{author}{\bibinfo{person}{Philip~M Podsakoff},
  \bibinfo{person}{Scott~B MacKenzie}, \bibinfo{person}{Jeong-Yeon Lee}, {and}
  \bibinfo{person}{Nathan~P Podsakoff}.} \bibinfo{year}{2003}\natexlab{}.
\newblock \showarticletitle{Common method biases in behavioral research: a
  critical review of the literature and recommended remedies.}
\newblock \bibinfo{journal}{\emph{Journal of Applied Psychology}}
  \bibinfo{volume}{88}, \bibinfo{number}{5} (\bibinfo{year}{2003}),
  \bibinfo{pages}{879}.
\newblock


\bibitem[Pratt et~al\mbox{.}(2020)]%
        {pratt2020editorial}
\bibfield{author}{\bibinfo{person}{Michael~G Pratt}, \bibinfo{person}{Sarah
  Kaplan}, {and} \bibinfo{person}{Richard Whittington}.}
  \bibinfo{year}{2020}\natexlab{}.
\newblock \showarticletitle{The tumult over transparency: Decoupling
  transparency from replication in establishing trustworthy qualitative
  research}.
\newblock \bibinfo{journal}{\emph{Administrative Science Quarterly}}
  \bibinfo{volume}{65}, \bibinfo{number}{1} (\bibinfo{year}{2020}),
  \bibinfo{pages}{1--19}.
\newblock


\bibitem[Pugh(2013)]%
        {pugh2013good}
\bibfield{author}{\bibinfo{person}{Allison~J Pugh}.}
  \bibinfo{year}{2013}\natexlab{}.
\newblock \showarticletitle{What good are interviews for thinking about
  culture? Demystifying interpretive analysis}.
\newblock \bibinfo{journal}{\emph{American Journal of Cultural Sociology}}
  \bibinfo{volume}{1} (\bibinfo{year}{2013}), \bibinfo{pages}{42--68}.
\newblock


\bibitem[Rhoads(2020)]%
        {rhoads2020whales}
\bibfield{author}{\bibinfo{person}{Robert~A Rhoads}.}
  \bibinfo{year}{2020}\natexlab{}.
\newblock \showarticletitle{“Whales tales” on the run: Anonymizing
  ethnographic data in an age of openness}.
\newblock \bibinfo{journal}{\emph{Cultural Studies and Critical Methodologies}}
  \bibinfo{volume}{20}, \bibinfo{number}{5} (\bibinfo{year}{2020}),
  \bibinfo{pages}{402--413}.
\newblock


\bibitem[Roose(2022)]%
        {rooseweb3}
\bibfield{author}{\bibinfo{person}{Kevin Roose}.}
  \bibinfo{year}{2022}\natexlab{}.
\newblock \showarticletitle{What is Web3?}
\newblock \bibinfo{journal}{\emph{The New York Times}} (\bibinfo{date}{Mar}
  \bibinfo{year}{2022}).
\newblock
\urldef\tempurl%
\url{https://www.nytimes.com/interactive/2022/03/18/technology/web3-definition-internet.html}
\showURL{%
\tempurl}


\bibitem[Scheuerman et~al\mbox{.}(2021)]%
        {scheuerman2021datasets}
\bibfield{author}{\bibinfo{person}{Morgan~Klaus Scheuerman},
  \bibinfo{person}{Alex Hanna}, {and} \bibinfo{person}{Emily Denton}.}
  \bibinfo{year}{2021}\natexlab{}.
\newblock \showarticletitle{Do datasets have politics? Disciplinary values in
  computer vision dataset development}.
\newblock \bibinfo{journal}{\emph{Proceedings of the ACM on Human-Computer
  Interaction}} \bibinfo{volume}{5}, \bibinfo{number}{CSCW2}
  (\bibinfo{year}{2021}), \bibinfo{pages}{1--37}.
\newblock


\bibitem[Scott(2013)]%
        {scott2013institutions}
\bibfield{author}{\bibinfo{person}{W~Richard Scott}.}
  \bibinfo{year}{2013}\natexlab{}.
\newblock \bibinfo{booktitle}{\emph{Institutions and organizations: Ideas,
  interests, and identities}}.
\newblock \bibinfo{publisher}{Sage publications}.
\newblock


\bibitem[Semenzin and Gandini(2021)]%
        {semenzin2021automating}
\bibfield{author}{\bibinfo{person}{Silvia Semenzin} {and}
  \bibinfo{person}{Alessandro Gandini}.} \bibinfo{year}{2021}\natexlab{}.
\newblock \showarticletitle{Automating trust with the blockchain? A critical
  investigation of “Blockchain 2.0” cultures}.
\newblock \bibinfo{journal}{\emph{Global Perspectives}} \bibinfo{volume}{2},
  \bibinfo{number}{1} (\bibinfo{year}{2021}), \bibinfo{pages}{24912}.
\newblock


\bibitem[Shin(2022)]%
        {shin2022}
\bibfield{author}{\bibinfo{person}{Laura Shin}.}
  \bibinfo{year}{2022}\natexlab{}.
\newblock \bibinfo{booktitle}{\emph{The cryptopians: Idealism, greed, lies, and
  the making of the first big cryptocurrency craze}}.
\newblock \bibinfo{publisher}{Hachette Book Group}, \bibinfo{address}{New
  York}.
\newblock


\bibitem[Shrestha(2019)]%
        {shrestha2019sov}
\bibfield{author}{\bibinfo{person}{Ajay~Kumar Shrestha}.}
  \bibinfo{year}{2019}\natexlab{}.
\newblock \bibinfo{title}{Stablecoins are doomed to fail}.
\newblock
  \bibinfo{howpublished}{\url{https://theconversation.com/transparency-and-privacy-empowering-people-through-blockchain-104887}}.
\newblock


\bibitem[Simon(1957)]%
        {simon1957}
\bibfield{author}{\bibinfo{person}{Herbert~A Simon}.}
  \bibinfo{year}{1957}\natexlab{}.
\newblock \bibinfo{booktitle}{\emph{Models of man: Social and rational}}.
\newblock \bibinfo{publisher}{Wiley}.
\newblock


\bibitem[Skinner(2011)]%
        {skinner2011social}
\bibfield{author}{\bibinfo{person}{Julia Skinner}.}
  \bibinfo{year}{2011}\natexlab{}.
\newblock \showarticletitle{Social media and revolution: The Arab spring and
  the occupy movement as seen through three information studies paradigms}.
\newblock \bibinfo{journal}{\emph{All Sprouts Content}}  \bibinfo{volume}{483}
  (\bibinfo{year}{2011}).
\newblock


\bibitem[Souleles(2018)]%
        {souleles2018study}
\bibfield{author}{\bibinfo{person}{Daniel Souleles}.}
  \bibinfo{year}{2018}\natexlab{}.
\newblock \showarticletitle{How to study people who do not want to be studied:
  Practical reflections on studying up}.
\newblock \bibinfo{journal}{\emph{PoLAR: Political and Legal Anthropology
  Review}} \bibinfo{volume}{41}, \bibinfo{number}{S1} (\bibinfo{year}{2018}),
  \bibinfo{pages}{51--68}.
\newblock


\bibitem[Stangor(2014)]%
        {stangor2014research}
\bibfield{author}{\bibinfo{person}{Charles Stangor}.}
  \bibinfo{year}{2014}\natexlab{}.
\newblock \bibinfo{booktitle}{\emph{Research methods for the behavioral
  sciences}}.
\newblock \bibinfo{publisher}{Cengage Learning}.
\newblock


\bibitem[Strauss and Corbin(1990)]%
        {strauss1990basics}
\bibfield{author}{\bibinfo{person}{Anselm Strauss} {and}
  \bibinfo{person}{Juliet Corbin}.} \bibinfo{year}{1990}\natexlab{}.
\newblock \bibinfo{booktitle}{\emph{Basics of qualitative research}}.
\newblock \bibinfo{publisher}{Sage Publications}.
\newblock


\bibitem[Swartz(2016)]%
        {swartz2016}
\bibfield{author}{\bibinfo{person}{Lana Swartz}.}
  \bibinfo{year}{2016}\natexlab{}.
\newblock \bibinfo{booktitle}{\emph{Another economy is possible}}.
\newblock \bibinfo{publisher}{Polity Malden}, Chapter Blockchain dreams:
  Imagining techno-economic alternatives after Bitcoin,
  \bibinfo{pages}{82--105}.
\newblock


\bibitem[Swartz(2018)]%
        {swartz2018bitcoin}
\bibfield{author}{\bibinfo{person}{Lana Swartz}.}
  \bibinfo{year}{2018}\natexlab{}.
\newblock \showarticletitle{What was Bitcoin, what will it be? The
  techno-economic imaginaries of a new money technology}.
\newblock \bibinfo{journal}{\emph{Cultural Studies}} \bibinfo{volume}{32},
  \bibinfo{number}{4} (\bibinfo{year}{2018}), \bibinfo{pages}{623--650}.
\newblock


\bibitem[Swartz(2022)]%
        {swartz2022theorizing}
\bibfield{author}{\bibinfo{person}{Lana Swartz}.}
  \bibinfo{year}{2022}\natexlab{}.
\newblock \showarticletitle{Theorizing the 2017 blockchain ICO bubble as a
  network scam}.
\newblock \bibinfo{journal}{\emph{New Media \& Society}} \bibinfo{volume}{24},
  \bibinfo{number}{7} (\bibinfo{year}{2022}), \bibinfo{pages}{1695--1713}.
\newblock


\bibitem[Tavory and Timmermans(2014)]%
        {tavory2014abductive}
\bibfield{author}{\bibinfo{person}{Iddo Tavory} {and} \bibinfo{person}{Stefan
  Timmermans}.} \bibinfo{year}{2014}\natexlab{}.
\newblock \bibinfo{booktitle}{\emph{Abductive analysis: Theorizing Qualitative
  Research}}.
\newblock \bibinfo{publisher}{University of Chicago Press}.
\newblock


\bibitem[Timmermans and Tavory(2012)]%
        {timmermans2012theory}
\bibfield{author}{\bibinfo{person}{Stefan Timmermans} {and}
  \bibinfo{person}{Iddo Tavory}.} \bibinfo{year}{2012}\natexlab{}.
\newblock \showarticletitle{Theory construction in qualitative research: From
  grounded theory to abductive analysis}.
\newblock \bibinfo{journal}{\emph{Sociological Theory}} \bibinfo{volume}{30},
  \bibinfo{number}{3} (\bibinfo{year}{2012}), \bibinfo{pages}{167--186}.
\newblock


\bibitem[Twenge et~al\mbox{.}(2022)]%
        {twenge2022specification}
\bibfield{author}{\bibinfo{person}{Jean~M Twenge}, \bibinfo{person}{Jonathan
  Haidt}, \bibinfo{person}{Jimmy Lozano}, {and} \bibinfo{person}{Kevin~M
  Cummins}.} \bibinfo{year}{2022}\natexlab{}.
\newblock \showarticletitle{Specification curve analysis shows that social
  media use is linked to poor mental health, especially among girls}.
\newblock \bibinfo{journal}{\emph{Acta psychologica}}  \bibinfo{volume}{224}
  (\bibinfo{year}{2022}), \bibinfo{pages}{103512}.
\newblock


\bibitem[Vaisey(2009)]%
        {vaisey2009motivation}
\bibfield{author}{\bibinfo{person}{Stephen Vaisey}.}
  \bibinfo{year}{2009}\natexlab{}.
\newblock \showarticletitle{Motivation and justification: A dual-process model
  of culture in action}.
\newblock \bibinfo{journal}{\emph{Amer. J. Sociology}} \bibinfo{volume}{114},
  \bibinfo{number}{6} (\bibinfo{year}{2009}), \bibinfo{pages}{1675--1715}.
\newblock


\bibitem[Voskobojnikov et~al\mbox{.}(2020)]%
        {voskobojnikov2020surviving}
\bibfield{author}{\bibinfo{person}{Artemij Voskobojnikov},
  \bibinfo{person}{Borke Obada-Obieh}, \bibinfo{person}{Yue Huang}, {and}
  \bibinfo{person}{Konstantin Beznosov}.} \bibinfo{year}{2020}\natexlab{}.
\newblock \showarticletitle{Surviving the cryptojungle: Perception and
  management of risk among North American cryptocurrency (non) users}. In
  \bibinfo{booktitle}{\emph{Financial Cryptography and Data Security: 24th
  International Conference, FC 2020, Kota Kinabalu, Malaysia, February 10--14,
  2020 Revised Selected Papers}}. Springer, \bibinfo{pages}{595--614}.
\newblock


\bibitem[Voskobojnikov et~al\mbox{.}(2021)]%
        {voskobojnikov2021u}
\bibfield{author}{\bibinfo{person}{Artemij Voskobojnikov},
  \bibinfo{person}{Oliver Wiese}, \bibinfo{person}{Masoud Mehrabi~Koushki},
  \bibinfo{person}{Volker Roth}, {and} \bibinfo{person}{Konstantin Beznosov}.}
  \bibinfo{year}{2021}\natexlab{}.
\newblock \showarticletitle{The U in crypto stands for usable: An empirical
  study of user experience with mobile cryptocurrency wallets}. In
  \bibinfo{booktitle}{\emph{Proceedings of the 2021 CHI Conference on Human
  Factors in Computing Systems}}. \bibinfo{pages}{1--14}.
\newblock


\bibitem[Walch(2019)]%
        {walch2019deconstructing}
\bibfield{author}{\bibinfo{person}{Angela Walch}.}
  \bibinfo{year}{2019}\natexlab{}.
\newblock \showarticletitle{Crypto assets: Legal and monetary perspectives}.
\newblock \bibinfo{publisher}{Oxford University Press}, Chapter Deconstructing
  'decentralization': Exploring the core claim of crypto systems,
  \bibinfo{pages}{39--68}.
\newblock


\bibitem[Walden(2020)]%
        {walden2020progressive}
\bibfield{author}{\bibinfo{person}{Jesse Walden}.}
  \bibinfo{year}{2020}\natexlab{}.
\newblock \bibinfo{title}{Progressive decentralization: A playbook for building
  crypto applications}.
\newblock
  \bibinfo{howpublished}{\url{https://a16z.com/2020/01/09/progressive-decentralization-crypto-product-management/}}.
\newblock
\newblock
\shownote{Accessed: 2022-11-01}.


\bibitem[White(2023)]%
        {whiteweb3goinggreat}
\bibfield{author}{\bibinfo{person}{Molly White}.}
  \bibinfo{year}{2023}\natexlab{}.
\newblock \bibinfo{title}{Web3 is Going Great}.
\newblock \bibinfo{howpublished}{\url{https://web3isgoinggreat.com/}}.
\newblock
\newblock
\shownote{Accessed: 2023-06-20}.


\bibitem[Winecoff and Watkins(2022)]%
        {winecoff2022artificial}
\bibfield{author}{\bibinfo{person}{Amy~A Winecoff} {and}
  \bibinfo{person}{Elizabeth~A Watkins}.} \bibinfo{year}{2022}\natexlab{}.
\newblock \showarticletitle{Artificial concepts of artificial intelligence:
  Institutional compliance and resistance in AI startups}. In
  \bibinfo{booktitle}{\emph{Proceedings of the 2022 AAAI/ACM Conference on AI,
  Ethics, and Society (AIES’22)}}.
\newblock


\bibitem[Wood(2018)]%
        {wood2018}
\bibfield{author}{\bibinfo{person}{Gavin Wood}.}
  \bibinfo{year}{2018}\natexlab{}.
\newblock \bibinfo{title}{Why We Need Web 3.0}.
\newblock
\newblock
\urldef\tempurl%
\url{https://gavofyork.medium.com/why-we-need-web-3-0-5da4f2bf95ab}
\showURL{%
\tempurl}


\bibitem[Wuthnow(2011)]%
        {wuthnow2011taking}
\bibfield{author}{\bibinfo{person}{Robert~J Wuthnow}.}
  \bibinfo{year}{2011}\natexlab{}.
\newblock \showarticletitle{Taking talk seriously: Religious discourse as
  social practice}.
\newblock \bibinfo{journal}{\emph{Journal for the Scientific Study of
  Religion}} \bibinfo{volume}{50}, \bibinfo{number}{1} (\bibinfo{year}{2011}),
  \bibinfo{pages}{1--21}.
\newblock


\end{thebibliography}
